\documentclass{aa}
\usepackage{natbib}
\bibpunct{(}{)}{;}{a}{}{,} % to follow the A&A style
\usepackage{graphics}   % standard graphics specifications
\usepackage{graphicx}   % alternative graphics specificati.jpgons
\usepackage{epstopdf}
\usepackage{longtable}   % helps with long table options
\usepackage{url}      % for on-line citations
\usepackage{bm}        % special 'bold-math' package
\usepackage[varg]{txfonts}
\usepackage{pdflscape}
\usepackage{supertabular}
\usepackage{hyperref}   %link
\hypersetup{draft}
\usepackage{rotating}

\newcommand{\msunyr}{\ensuremath{M_{\odot}{\rm yr}^{-1}}}   % msun/yr
\newcommand{\msun}{\ensuremath{M_{\odot}}}   % msun
\newcommand{\mini}{\ensuremath{M_{\rm ini}}}                         %M ini
\newcommand{\rsun}{\ensuremath{\mathit{R}_{\odot}}}                  % solar radius
\newcommand{\zsun}{\ensuremath{\mathit{Z}_{\odot}}}                  % solar metal content

\newcommand{\mdot}{\ensuremath{\dot{M}}}                             % mass loss rate
\newcommand{\teff}{\ensuremath{\mathit{T}_{\rm eff}}}                % effectieve temperatuur
\newcommand{\K}{\ensuremath{\mathrm{K}}}                 % stellar T
\newcommand{\vrot}{\ensuremath{\upsilon_{\rm rot}}}                         %rotational velocity
\newcommand{\vcrit}{\ensuremath{\upsilon_{\rm crit}}}                         %rotational velocity
\newcommand{\eddesc}{\ensuremath{\Gamma_{\mathrm{es}}}}                 % optical depth of the Lyman continuum
\newcommand{\izw}{I\,Zw\,18}   

\hypersetup{
    bookmarks=true,         % show bookmarks bar?
    unicode=false,          % non-Latin characters in Acrobat?s bookmarks
    colorlinks=true,       % false: boxed links; true: colored links
    linkcolor=blue,          % color of internal links (change box color with linkbordercolor)
    citecolor=blue,        % color of links to bibliography
    filecolor=blue,      % color of file links
    urlcolor=blue           % color of external links
}
\begin{document}

 \title{Grids of stellar models with rotation}

   \subtitle{\boldmath IV. Models from $1.7$ to $120\,M_\odot$ at a metallicity $Z = 0.0004$}
   
   \titlerunning{Grids of stellar models with rotation. IV}

   \author{J.~H. Groh \inst{1}
          \and
          S.~Ekstr\"om \inst{2}
          \and
          C.~Georgy \inst{2}          
          \and
          G.~Meynet \inst{2}
          \and
          A.~Choplin \inst{2} 
          \and
          P.~Eggenberger \inst{2} 
           \and
          R.~Hirschi \inst{3}                    
          \and
          A.~Maeder \inst{2}
          \and
          L.~J.~Murphy \inst{1}
          \and
          I.~Boian \inst{1}
          \and
          E. J. Farrell \inst{1}
          }

   \institute{
           $^1$School of Physics, Trinity College Dublin, the University of Dublin, Dublin, Ireland;  \email{jose.groh@tcd.ie} \\
            $^2$Department of Astronomy, University of Geneva, Maillettes 51, 1290 Versoix, Switzerland\\
            $^3$Astrophysics Group, Keele University, Keele, Staffordshire, ST5 5BG, UK\\
             }

   \authorrunning{Groh et al.}
 
   \date{Received ; accepted}
 
  \abstract{The effects of rotation on stellar evolution are particularly important at low metallicity, when mass loss by stellar winds diminishes and the surface enrichment due to rotational mixing becomes relatively more pronounced than at high metallicities. Here we investigate the impact of rotation and metallicity on stellar evolution. Using a similar physics as in our previous large grids of models at $Z=0.002$ and $Z=0.014$, we compute stellar evolution models with the Geneva code for rotating and nonrotating stars with initial masses (\mini) between 1.7 and 120~\msun\ and $Z=0.0004$ (1/35 solar). This is comparable to the metallicities of the most metal poor galaxies observed so far, such as \izw. Concerning massive stars, both rotating and nonrotating models spend most of their core-helium burning phase with an effective temperature higher than 8000~\K.  Stars become red supergiants only at the end of their lifetimes, and few red supergiants are expected. Our models predict very few to no classical Wolf-Rayet stars as a results of weak stellar winds at low metallicity. The most massive stars end their lifetimes as luminous blue supergiants or luminous blue variables, a feature that is not predicted by models with higher initial metallicities. Interestingly, due to the behavior of the intermediate convective zone, the mass domain of stars producing pair-instability supernovae is smaller at Z=0.0004 than at Z=0.002. We find that during the main sequence (MS) phase, the ratio between nitrogen and carbon abundances (N/C) remains unchanged for nonrotating models. However, N/C increases by factors of 10-20 in rotating models at the end of the MS. Cepheids coming from stars with  $\mini > 4-6~\msun$ are beyond the core helium burning phase and spend little time in the instability strip. Since they would evolve towards cooler effective temperatures, these Cepheids should show an increase of the pulsation period as a function of age. }
\maketitle
   
%=================================================================================
% INTRODUCTION
%=================================================================================
\section{Introduction}

%why evolution at low Z is interesting?
Stellar evolution at low metallicity has a significant impact in many topics of astrophysics, such as the photometric and chemical evolution of metal-poor galaxies at different redshifts \citep[e.g.,][]{tolstoy09,stark16}, 
the observable properties of integrated stellar populations  \citep[e.g.,][]{Vazdekis2010,eldridge17}, the nature of supernovae (SN) and gamma-ray burst progenitors \citep[e.g.,][]{modjaz08,Japelj16,Schulze2016}, 
the amount of ionizing flux at high redshift \citep{Levesque2012,goetberg17,goetberg18}, nucleosynthesis \citep[e.g.,][]{Chiappini2011}, and the rates of gravitational-wave signals from merging black holes \citep[BH;][]{LIGOII2016,abbott16a,abbott16b,LIGOI2016,abbott17a,abbott17b,belc16} and neutron star (NS) systems \citep{abbott17c}. Current observational facilities have obtained an increasing amount of observational data for low-metallicity environments, and future missions will provide an unprecedented view on those systems. 

Investigating stellar evolution at metallicities similar to those of the most metal-poor nearby galaxies such as \izw\ \citep{Zw1966, Searle1972, Izo1999} is of particular interest. These galaxies have metallicities as low as $\sim1/35$ of the solar value (\zsun), offering a unique opportunity to probe the impact of a metal-poor environment on the evolution of stars and the properties of a recent starburst episode \citep{Izo2004}. In addition, these blue-compact dwarf galaxies bridge the gap between nearby systems where individual stars can be observed and metal-poor, high-redshift galaxies for which only integrated light of stellar populations can be detected. Recent studies have presented surprising results showing that the ultraviolet emission from distant, low-metallicity stellar populations is often not reproduced by current models \citep[e.g.,][]{steidel14, steidel16,stark15}. This highlights the challenges in understanding stellar evolution at low metallicities, and ultimately exposes the limitations of the current stellar evolution models that are nevertheless required for fully extracting physical information from the observations of stars and stellar populations. 

There have been several recent efforts to produce grids of stellar models for stars with metallicities similar to those of local, metal-poor dwarf galaxies such as \izw. These grids have different physical assumptions, in particular regarding the implementation of rotation and internal magnetic fields. Nonrotating models have been computed by the Padova group \citep{Bressan2012}, while \citet{Choi2016} presented an extensive grid of MESA stellar evolution models \citep{paxton11,paxton13,paxton15} that include rotation but no internal magnetic fields. \citet{SZ2015} used the Bonn stellar evolution code and considered rotating models including the effects of an internal magnetic field amplified in differentially rotating radiative layers \citep{Spruit2002}. Both Bonn and MESA models use a diffusive approach to treat the transport of angular momentum in the stellar interior.  

%Fig HR Diagrams
\begin{figure*}
\includegraphics[width=0.47\textwidth]{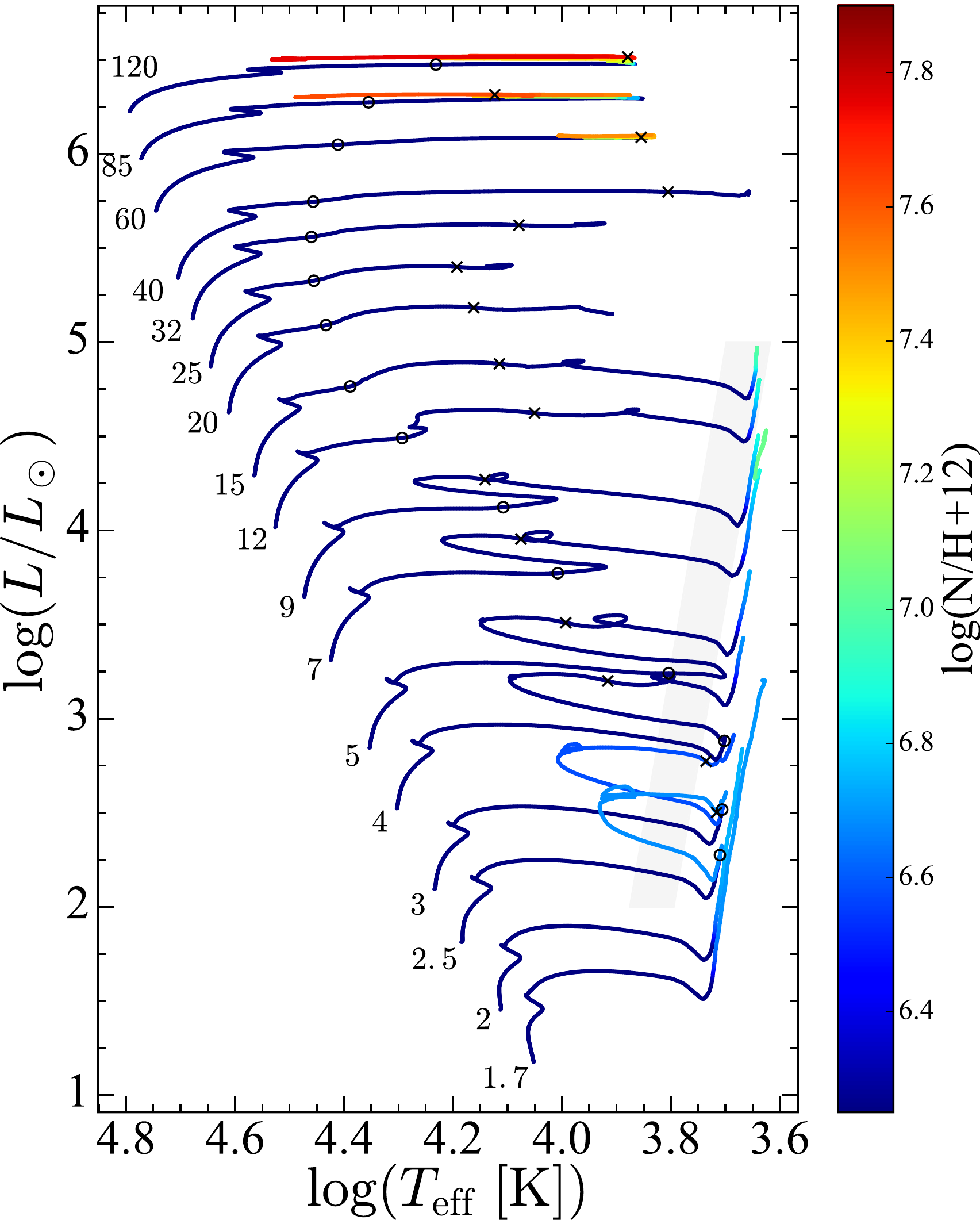}\hfill\includegraphics[width=0.47\textwidth]{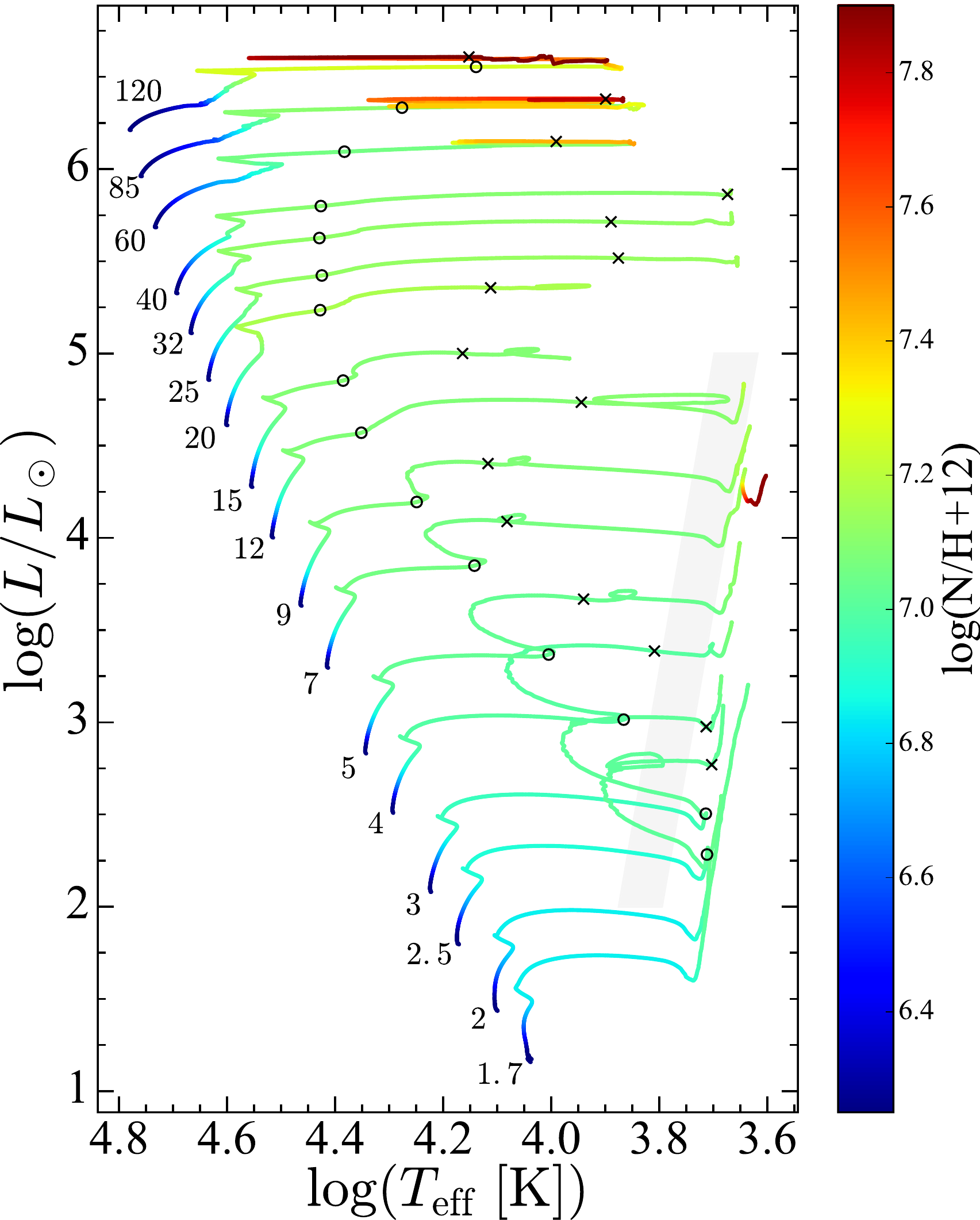}
\caption{Evolutionary tracks in the theoretical HR diagram for nonrotating (left panel) and rotating models (right panel). Each track is labeled with its initial mass. The beginning (when 0.003 in mass fraction of helium has been burnt at the center) and end of the core He-burning phase are indicated
by open circles and crosses, respectively. The Cepheid instability strip is indicated by the light gray region \citep{Tammann2003}. The color scale indicates the surface number abundance of nitrogen on a logarithmic scale where the abundance of hydrogen is 12.
\label{Fig:HRDgen}}
\end{figure*}

In this paper we present a new grid of stellar evolution models computed with the Geneva code for the mass range of 1.7 to 120 $M_\odot$ and  metallicity $Z$=0.0004, with a similar set of physical ingredients as in our two previous grids of models for $Z$=0.014 \citep{Ekstrom2012a} and $Z=0.002$ \citep{Georgy2013a}. Our treatment of rotation is different from those of \citet{SZ2015} and \citet{Choi2016}, offering an original view on the effects of rotation at low metallicity when the assumption of ``shellular'' rotations is adopted \citep{Zahn1992}. 

This paper is organized as follows. The physical ingredients of the numerical simulations and the electronic tables are described
in Sect.~\ref{Sec:inputs}. Section~\ref{Sec:properties} discusses the main properties of the models. The effects of metallicity are discussed in Sect.~\ref{Sec:metal}, and the impact of the various angular momentum processes are presented in Sect.~\ref{Sec:comp}. We discuss the implications of our models for interpreting the galaxy  \izw\ in Sect.~\ref{Sec:izw18}, and present our conclusions and future perspectives in Sect.~\ref{Sec:conc}.

%=================================================================================
% PHYSICS INPUTS
%=================================================================================
\section{Stellar models: physical ingredients and electronic tables\label{Sec:inputs}}

\defcitealias{Ekstrom2012a}{Papers I}
\defcitealias{georgy12a}{ II}
\defcitealias{Georgy2013a}{ III}

The physical inputs of the models presented in this paper are the same as those from \citet[hereafter Paper I]{Ekstrom2012a}, \citet[hereafter Paper II]{georgy12a},  and \citet[hereafter Paper III]{Georgy2013a}.  We refer the reader to those papers for a detailed description of the physical inputs, and we briefly outline the main ingredients. The initial abundances of H, He, and metals are set to ${\rm X} = 0.7507$, ${\rm Y} = 0.2489$, and ${\rm Z} = 0.0004$. The mixture of heavy elements follows \citetalias{Ekstrom2012a}, downscaled to a metallicity of ${\rm Z} = 0.0004$.
The convective cores during the H- and He-burning phases are extended by including an overshooting parameter equal to $0.1 H_\mathit{P}$ , where $H_\mathit{P}$ is the pressure scale-height scale at the Schwarzschild convective boundary. The recipes for mass-loss rates ($\mdot$) and their metallicity dependence are taken as in \citetalias{Ekstrom2012a} and \citetalias{Georgy2013a}. We scale the mass-loss rates with metallicity as  $\mdot (Z) = (Z/Z\sun)^\alpha \mdot(Z\sun)$. For the MS and blue supergiant phases, we assume $\alpha=0.85$ or 0.50 when the \citet{Vink2001} or \citet{dejager88} recipes are used, respectively. For the Wolf-Rayet (WR) phase, we assume $\alpha=0.66$, following \citet{eldridgevink06}. For other phases, such as when the effective temperature $\teff$ is lower than $\log (\teff / K = 3.7)$, no metallicity scaling is applied. The physics of rotation is the same as in our two grids of stellar models.

%=================================================================================
% STELLAR MODELS AND ELECTRONIC TABLES
%=================================================================================

We computed stellar evolution models for the following initial masses: $1.7$, $2$, $2.5$, $3$, $4$, $5$, $7$, $9$, $12$, $15$, $20$, $25$, $32$, $40$, $60$, $85$, and $120\,M_\odot$. For each mass, we computed both a nonrotating and a rotating model with a ratio between the equatorial surface rotational velocity (\vrot) and critical rotational velocity (\vcrit) of 0.4 at the zero-age main sequence (ZAMS). The models are evolved up to the end of core carbon burning ($M_{\rm ini} \ge 12\,M_\odot$), the early asymptotic giant branch ($2.5\,M_\odot \le M_{\rm ini} \le 9\,M_\odot$), or the helium flash ($M_{\rm ini} \le 2\,M_\odot$).

Similarly to \citetalias{Ekstrom2012a} and \citetalias{Georgy2013a}, electronic tables of the evolutionary sequences are publicly available\footnote{See \href{http://obswww.unige.ch/Recherche/evol/-Database-}{http://obswww.unige.ch/Recherche/evol/-Database-}\\or the CDS database at \href{http://vizier.u-strasbg.fr/viz-bin/VizieR-2}{http://vizier.u-strasbg.fr/viz-bin/VizieR-2}.}. For each model, the evolutionary track is described by 400 selected data points, with each one corresponding to a given evolutionary stage. Points of different evolutionary tracks with the same number correspond to similar stages to facilitate the interpolation of evolutionary tracks. The points are numbered as described in \citet{Ekstrom2012a}. The new grid can thus be used as input for computing interpolated tracks, isochrones, and population synthesis models using the publicly available Geneva tools\footnote{\href{https://obswww.unige.ch/Recherche/evoldb/index/}{https://obswww.unige.ch/Recherche/evoldb/index/}}.
A detailed description of the online tools is presented in \citet{Georgy2014}.

%=================================================================================
% PROPERTIES OF THE MODELS
%=================================================================================
\section{Properties of the stellar models \label{Sec:properties}}

%\afterpage{\clearpage}
In this section we describe the main features of the grid of stellar evolution models at $Z=0.0004$. Table~\ref{Tab:properties} presents the main properties of the models at the end of each core burning phase.  We discuss some of the differences between our models compared to MESA \citep{Choi2016} and Bonn models \citep{SZ2015} in Sect. 5.

\begin{sidewaystable*}
\begin{minipage}{\textwidth}
\caption{Properties of the $Z=0.0004$ stellar models at the end of the H-, He-, and C-burning phases. Columns 1 to 4 give the initial mass of the model, the initial ratio between the rotational velocity at the equator and the critical rotational velocity, the initial rotational velocity at the equator, and the time-averaged equatorial surface velocity during the MS phase,  respectively. Columns 5 to 11 show properties of the stellar models at the end of the core H-burning phase: age, actual mass, surface equatorial velocity, ratio of the equatorial surface velocity to the critical velocity, mass fraction of helium at the surface, ratios in mass fraction of the nitrogen to carbon abundances at the surface, and of nitrogen to oxygen at the surface,  respectively.  Columns 12 to 19 and 20 to 27 show properties of the models at the end of the core He- and C-burning phases, 
respectively. The columns labeled $M$, $V_{\rm eq}$, $Y_{\rm surf}$, N/C and N/O have the same meaning as the corresponding columns for the end of the core hydrogen burning phase. The quantity $\tau_{\rm He}$ ($\tau_{\rm C}$) corresponds to the duration of the He- (C-) core burning phase, $P_{\rm rot}$ is the spin period of the star at the surface, $\Omega_{\rm C}/\Omega_{\rm S}$ is the ratio of the angular velocity of the core to that of the surface. }\label{Tab:properties}
\centering
\scalebox{0.6}{%
\begin{tabular}{rrrr|rrrrrrr|rrrrrrrr|rrrrrrrr}
\hline\hline
\multicolumn{4}{c|}{} & \multicolumn{7}{c|}{End of H-burning} & \multicolumn{8}{c|}{End of He-burning} & \multicolumn{8}{c}{End of C-burning}\\
$M_{\rm ini}$ & $\vrot/\vcrit$ & $V_{\rm eq}$ & $\bar{V}_{\rm MS}$ & $\tau_{\rm H}$ & $M$ & $V_{\rm eq}$ & $V_{\rm eq}/V_{\rm crit}$ & $Y_{\rm surf}$ & ${\rm N}/{\rm C}$ & ${\rm N}/{\rm O}$ & $\tau_{\rm He}$ & $M$ & $V_{\rm eq}$ & $P_{\rm rot}$ & $\Omega_{\rm C}/\Omega_{\rm S}$ & $Y_{\rm surf}$ & ${\rm N}/{\rm C}$ & ${\rm N}/{\rm O}$ & $\tau_{\rm C}$ & $M$ & $V_{\rm eq}$ & $P_{\rm rot}$ & $\Omega_{\rm C}/\Omega_{\rm S}$ & $Y_{\rm surf}$ & ${\rm N}/{\rm C}$ & ${\rm N}/{\rm O}$ \\
$M_{\sun}$ & & \multicolumn{2}{c|}{km s$^{-1}$} & Myr & $M_{\sun}$ & km s$^{-1}$ & & \multicolumn{3}{c|}{mass fract.} & Myr & $M_{\sun}$ & km s$^{-1}$ & day & & \multicolumn{3}{c|}{mass fract.} & kyr & $M_{\sun}$ & km s$^{-1}$ & day & & \multicolumn{3}{c}{mass fract.}\\
\hline
$120.00$ & $0.00$ & $  0.$ &   -- & $    2.606$ & $118.25$ &   -- & -- & $0.2489$ & $  0.2885$ & $  0.1152$ & $    0.269$ & $ 92.52$ &   -- & -- & -- & $0.6585$ & $ 69.3650$ & $ 59.8911$ & $    0.021$ & $ 87.49$ &  -- & -- & -- & $0.7042$ & $ 66.2987$ & $116.4622$ \\
$120.00$ & $0.40$ & $463.$ & $501.$ & $    3.050$ & $115.92$ & $156.$ & $0.231$ & $0.4238$ & $  5.5363$ & $  2.1020$ & $    0.270$ & $ 93.20$ & $  0.$ & $6.158\cdot10^{08}$ & $2.854\cdot10^{09}$ & $0.7973$ & $ 58.4043$ & $ 53.1908$ & $    0.002$ & $ 92.51$ & $  0.$ & $1.351\cdot10^{06}$ & $1.087\cdot10^{08}$ & $0.8012$ & $ 59.4490$ & $ 58.2812$ \\
$ 85.00$ & $0.00$ & $  0.$ & -- & $    2.975$ & $ 83.96$ & -- & -- & $0.2489$ & $  0.2885$ & $  0.1152$ & $    0.297$ & $ 66.15$ & -- & -- & -- & $0.6156$ & $ 44.5292$ & $ 15.2587$ & $    0.040$ & $ 66.04$ & -- & -- & -- & $0.6208$ & $ 49.8008$ & $ 16.8616$ \\
$ 85.00$ & $0.40$ & $469.$ & $470.$ & $    3.464$ & $ 82.86$ & $233.$ & $0.394$ & $0.3482$ & $  3.4577$ & $  1.2026$ & $    0.306$ & $ 61.30$ & $  2.$ & $2.783\cdot10^{04}$ & $5.702\cdot10^{04}$ & $0.7391$ & $ 43.5172$ & $ 19.3621$ & $    0.002$ & $ 57.81$ & $  0.$ & $4.534\cdot10^{13}$ & $1.931\cdot10^{15}$ & $0.7785$ & $ 63.5034$ & $ 47.5795$ \\
$ 60.00$ & $0.00$ & $  0.$ & -- & $    3.507$ & $ 59.43$ & -- & -- & $0.2489$ & $  0.2885$ & $  0.1152$ & $    0.337$ & $ 47.36$ & -- & -- & -- & $0.4738$ & $ 14.0549$ & $  3.6620$ & $    0.016$ & $ 45.70$ & -- & -- & -- & $0.5412$ & $ 26.0034$ & $  7.1750$ \\
$ 60.00$ & $0.40$ & $435.$ & $427.$ & $    4.046$ & $ 59.01$ & $195.$ & $0.328$ & $0.3226$ & $  3.5990$ & $  1.0319$ & $    0.343$ & $ 51.54$ & $  5.$ & $4.588\cdot10^{03}$ & $6.297\cdot10^{04}$ & $0.5369$ & $ 14.0556$ & $  3.4987$ & $    0.011$ & $ 49.91$ & $  1.$ & $6.666\cdot10^{04}$ & $3.753\cdot10^{07}$ & $0.5567$ & $ 15.8821$ & $  4.0306$ \\
$ 40.00$ & $0.00$ & $  0.$ & -- & $    4.468$ & $ 39.76$ & -- & -- & $0.2489$ & $  0.2885$ & $  0.1152$ & $    0.416$ & $ 39.46$ & -- & -- & -- & $0.2489$ & $  0.2885$ & $  0.1152$ & $    0.072$ & $ 37.83$ & -- & -- & -- & $0.2489$ & $  0.2892$ & $  0.1153$ \\
$ 40.00$ & $0.40$ & $393.$ & $328.$ & $    5.144$ & $ 39.66$ & $482.$ & $0.889$ & $0.3158$ & $  6.2331$ & $  1.1025$ & $    0.446$ & $ 37.03$ & $  1.$ & $8.406\cdot10^{04}$ & $2.269\cdot10^{06}$ & $0.3265$ & $  6.8653$ & $  1.1755$ & $    0.061$ & $ 34.58$ & $  0.$ & $3.120\cdot10^{06}$ & $3.292\cdot10^{09}$ & $0.3392$ & $  7.4605$ & $  1.2540$ \\
$ 32.00$ & $0.00$ & $  0.$ & -- & $    5.264$ & $ 31.87$ & -- & -- & $0.2489$ & $  0.2885$ & $  0.1152$ & $    0.501$ & $ 31.78$ & -- & -- & -- & $0.2489$ & $  0.2885$ & $  0.1152$ & $    0.144$ & $ 31.77$ & -- & -- & -- & $0.2489$ & $  0.2885$ & $  0.1152$ \\
$ 32.00$ & $0.40$ & $366.$ & $306.$ & $    6.174$ & $ 31.80$ & $371.$ & $0.606$ & $0.3130$ & $  9.1039$ & $  1.1439$ & $    0.542$ & $ 31.61$ & $  2.$ & $1.073\cdot10^{04}$ & $3.142\cdot10^{05}$ & $0.3148$ & $  9.2551$ & $  1.1550$ & $    0.124$ & $ 30.52$ & $  1.$ & $8.788\cdot10^{04}$ & $1.381\cdot10^{08}$ & $0.3241$ & $ 10.0297$ & $  1.2113$ \\
$ 25.00$ & $0.00$ & $  0.$ & -- & $    6.481$ & $ 24.94$ & -- & -- & $0.2489$ & $  0.2885$ & $  0.1152$ & $    0.637$ & $ 24.91$ & -- & -- & -- & $0.2489$ & $  0.2885$ & $  0.1152$ & $    0.380$ & $ 24.91$ & -- & -- & -- & $0.2489$ & $  0.2885$ & $  0.1152$ \\
$ 25.00$ & $0.40$ & $343.$ & $280.$ & $    7.683$ & $ 24.90$ & $310.$ & $0.565$ & $0.3096$ & $ 12.8248$ & $  1.1727$ & $    0.650$ & $ 24.79$ & $  6.$ & $2.850\cdot10^{03}$ & $8.852\cdot10^{04}$ & $0.3108$ & $ 13.0079$ & $  1.1805$ & $    0.242$ & $ 23.97$ & $  1.$ & $3.325\cdot10^{04}$ & $6.561\cdot10^{07}$ & $0.3241$ & $ 14.9144$ & $  1.2619$ \\
$ 20.00$ & $0.00$ & $  0.$ & -- & $    7.900$ & $ 19.97$ & -- & -- & $0.2489$ & $  0.2885$ & $  0.1152$ & $    0.858$ & $ 19.96$ & -- & -- & -- & $0.2489$ & $  0.2885$ & $  0.1152$ & $    0.743$ & $ 19.96$ & -- & -- & -- & $0.2489$ & $  0.2885$ & $  0.1152$ \\
$ 20.00$ & $0.40$ & $314.$ & $269.$ & $   10.216$ & $ 19.95$ & $322.$ & $0.568$ & $0.3280$ & $ 22.0435$ & $  1.3151$ & $    0.841$ & $ 19.90$ & $ 11.$ & $4.497\cdot10^{02}$ & $1.514\cdot10^{04}$ & $0.3296$ & $ 22.4326$ & $  1.3239$ & $    0.455$ & $ 19.90$ & $  0.$ & $3.636\cdot10^{15}$ & $9.760\cdot10^{18}$ & $0.3304$ & $ 22.6263$ & $  1.3283$ \\
$ 15.00$ & $0.00$ & $  0.$ & -- & $   10.949$ & $ 14.96$ & -- & -- & $0.2489$ & $  0.2885$ & $  0.1152$ & $    1.276$ & $ 14.94$ & -- & -- & -- & $0.2489$ & $  0.2885$ & $  0.1152$ & $    3.191$ & $ 14.92$ & -- & -- & -- & $0.3123$ & $  3.5286$ & $  0.7647$ \\
$ 15.00$ & $0.40$ & $319.$ & $242.$ & $   13.171$ & $ 14.95$ & $200.$ & $0.375$ & $0.2838$ & $ 14.4660$ & $  1.0445$ & $    1.411$ & $ 14.90$ & $ 25.$ & $1.012\cdot10^{02}$ & $4.312\cdot10^{03}$ & $0.2846$ & $ 14.7696$ & $  1.0515$ & $    1.828$ & $ 14.90$ & $  3.$ & $2.090\cdot10^{03}$ & $8.209\cdot10^{06}$ & $0.2857$ & $ 15.1674$ & $  1.0605$ \\
$ 12.00$ & $0.00$ & $  0.$ & -- & $   14.964$ & $ 11.98$ & -- & -- & $0.2489$ & $  0.2885$ & $  0.1152$ & $    1.744$ & $ 11.97$ & -- & -- & -- & $0.2489$ & $  0.2885$ & $  0.1152$ & $    6.527$ & $ 11.93$ & -- & -- & -- & $0.2797$ & $  2.9684$ & $  0.6025$ \\
$ 12.00$ & $0.40$ & $285.$ & $230.$ & $   17.822$ & $ 11.98$ & $235.$ & $0.450$ & $0.2800$ & $ 14.6114$ & $  1.0281$ & $    1.886$ & $ 11.96$ & $  9.$ & $5.753\cdot10^{02}$ & $2.483\cdot10^{04}$ & $0.2808$ & $ 14.9383$ & $  1.0353$ & $    5.890$ & $ 11.91$ & $  2.$ & $1.389\cdot10^{04}$ & $7.678\cdot10^{07}$ & $0.3403$ & $ 36.2088$ & $  1.4780$ \\
$  9.00$ & $0.00$ & $  0.$ & -- & $   24.985$ & $  9.00$ & -- & -- & $0.2489$ & $  0.2885$ & $  0.1152$ & $    2.796$ & $  8.99$ & -- & -- & -- & $0.2489$ & $  0.2885$ & $  0.1152$ & $    4.760$ & $  8.96$ & -- & -- & -- & $0.2603$ & $  2.7311$ & $  0.5473$ \\
$  9.00$ & $0.40$ & $270.$ & $215.$ & $   29.563$ & $  9.00$ & $217.$ & $0.431$ & $0.2755$ & $ 14.7022$ & $  1.0121$ & $    3.204$ & $  8.99$ & $ 34.$ & $4.728\cdot10^{01}$ & $2.391\cdot10^{03}$ & $0.2761$ & $ 14.9728$ & $  1.0177$ & $    3.569$ & $  8.96$ & $  3.$ & $4.759\cdot10^{03}$ & $4.213\cdot10^{07}$ & $0.3129$ & $ 34.7414$ & $  1.3241$ \\
\cline{20-27}
$  7.00$ & $0.00$ & $  0.$ & -- & $   38.682$ & $  7.00$ & -- & -- & $0.2489$ & $  0.2885$ & $  0.1152$ & $    5.030$ & $  7.00$ & -- & -- & -- & $0.2489$ & $  0.2885$ & $  0.1152$ & \\
$  7.00$ & $0.40$ & $275.$ & $205.$ & $   45.281$ & $  7.00$ & $204.$ & $0.421$ & $0.2706$ & $ 11.6227$ & $  0.9457$ & $    5.231$ & $  7.00$ & $ 34.$ & $3.854\cdot10^{01}$ & $2.203\cdot10^{03}$ & $0.2711$ & $ 11.8618$ & $  0.9519$ & \\
$  5.00$ & $0.00$ & $  0.$ & -- & $   72.988$ & $  5.00$ & -- & -- & $0.2489$ & $  0.2885$ & $  0.1152$ & $   11.835$ & $  4.98$ & -- & -- & -- & $0.2489$ & $  0.2885$ & $  0.1152$ & \\
$  5.00$ & $0.40$ & $264.$ & $193.$ & $   86.400$ & $  5.00$ & $190.$ & $0.415$ & $0.2657$ & $  8.1520$ & $  0.8631$ & $   11.459$ & $  5.00$ & $ 25.$ & $6.189\cdot10^{01}$ & $4.319\cdot10^{03}$ & $0.2664$ & $  8.4294$ & $  0.8733$ & \\
$  4.00$ & $0.00$ & $  0.$ & -- & $  115.094$ & $  4.00$ & -- & -- & $0.2489$ & $  0.2885$ & $  0.1152$ & $   21.848$ & $  3.98$ & -- & -- & -- & $0.2489$ & $  0.2974$ & $  0.1169$ & \\
$  4.00$ & $0.40$ & $233.$ & $186.$ & $  137.296$ & $  4.00$ & $183.$ & $0.417$ & $0.2633$ & $  6.1577$ & $  0.7842$ & $   20.427$ & $  4.00$ & $  9.$ & $2.162\cdot10^{02}$ & $1.649\cdot10^{04}$ & $0.2638$ & $  6.3173$ & $  0.7921$ & \\
$  3.00$ & $0.00$ & $  0.$ & -- & $  215.821$ & $  3.00$ & -- & -- & $0.2489$ & $  0.2885$ & $  0.1152$ & $   51.303$ & $  2.98$ & -- & -- & -- & $0.2500$ & $  0.8770$ & $  0.2452$ & \\
$  3.00$ & $0.40$ & $240.$ & $177.$ & $  258.008$ & $  3.00$ & $174.$ & $0.422$ & $0.2611$ & $  4.2777$ & $  0.6544$ & $   42.400$ & $  2.98$ & $  6.$ & $3.303\cdot10^{02}$ & $2.893\cdot10^{04}$ & $0.2720$ & $  7.1652$ & $  0.8027$ & \\
$  2.50$ & $0.00$ & $  0.$ & -- & $  329.121$ & $  2.50$ & -- & -- & $0.2489$ & $  0.2885$ & $  0.1152$ & $   90.633$ & $  2.48$ & -- & -- & -- & $0.2542$ & $  1.3759$ & $  0.3119$ & \\
$  2.50$ & $0.40$ & $211.$ & $172.$ & $  398.525$ & $  2.50$ & $169.$ & $0.429$ & $0.2605$ & $  3.3981$ & $  0.5558$ & $   74.790$ & $  2.47$ & $  6.$ & $2.800\cdot10^{02}$ & $2.650\cdot10^{04}$ & $0.2812$ & $  8.3793$ & $  0.7865$ & \\
\cline{12-19}
$  2.00$ & $0.00$ & $  0.$ & -- & $  582.601$ & $  2.00$ &-- & -- & $0.2489$ & $  0.2885$ & $  0.1152$ & \\
$  2.00$ & $0.40$ & $197.$ & $170.$ & $  716.745$ & $  2.00$ & $171.$ & $0.451$ & $0.2629$ & $  2.7703$ & $  0.4722$ & \\
$  1.70$ & $0.00$ & $  0.$ & -- & $  929.506$ & $  1.70$ & -- & -- & $0.2489$ & $  0.2885$ & $  0.1152$ & \\
$  1.70$ & $0.40$ & $185.$ & $163.$ & $ 1175.367$ & $  1.70$ & $172.$ & $0.453$ & $0.2707$ & $  2.8011$ & $  0.4685$ & \\
\cline{1-11}
\end{tabular}}
\end{minipage}
\end{sidewaystable*}

%=================================================================================
\subsection{Evolution in the HR diagram\label{Subsec:HRD}}

Figure \ref{Fig:HRDgen}  shows the evolutionary tracks for all  the models. The most important feature are the following:
\begin{itemize}
\item The width of the MS band increases in the high mass range. This was found in Papers I and III, and is a feature that is even more pronounced at low metallicity and is a result of weaker mass-loss rates by stellar winds and larger cores \citep{Meynet1994}.
\item  As expected, for a star of a given initial mass, rotating tracks are slightly bluer and more luminous than the corresponding nonrotating tracks. This is caused by the lower opacities and larger cores at low metallicity compared to high-metallicity models.
\item  The vast majority of the core He-burning phase occurs as a blue supergiant, at relatively high effective temperatures for massive stars. Stars only become red supergiants (RSG) after core-He burning. For instance, the core He-burning phase occurs at $\log \teff/\K >  3.90$ for masses larger than 4 $M_\odot$ for rotating models (6~\msun\ for nonrotating models).
\item For initial masses between 2.5 and 3-4 $M_\odot$ (depending on rotation), we find the presence of complete blue loops, that is, loops that start and finish at the location of the Hayashi track. The presence of blue loops has been discussed in detail by \citet{walmswell15}, who propose that the excess He and its associated high mean-molecular weight above the H-burning shell is the main reason behind the blue loops.
\item For initial masses above 3-5 $M_\odot$ (depending on rotation), the Cepheid instability strip is crossed after the end of the core He-burning phase. This implies a short lifetime in the instability strip and an evolution towards lower effective temperatures, that is, towards an increase of the pulsation period as a function of age.
\item There is significant diversity in the end points of the evolution of the massive stars, comprising RSGs, blue supergiants (BSG), yellow supergiants (YSG), and luminous blue variables (LBVs; see the bottom left panel of Fig.~\ref{Fig:end}). However, the models do not predict classical WR stars as the endpoints of the evolution.

\end{itemize}

\subsection{Internal structure and lifetimes\label{Subsec:time}}

\begin{figure}
\centering \includegraphics[width=0.89\columnwidth]{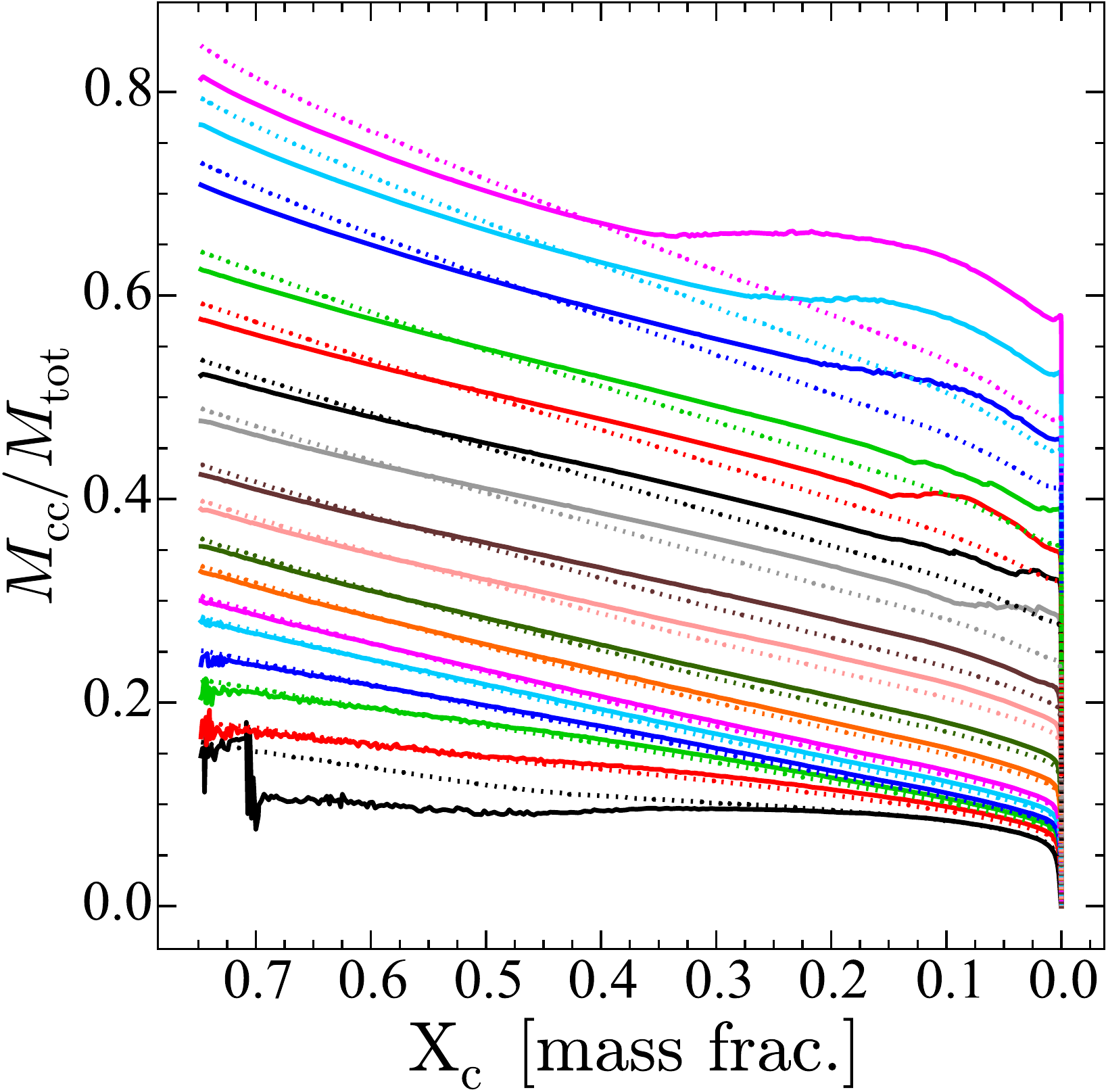}
\caption{Evolution of the fractional mass of the convective cores during the MS as a function of the mass fraction of hydrogen at the center. The bottom black line is for the  $1.7\,M_\odot$ model, and the top magenta line for the $120\,M_\odot$ model. The other intermediate lines show models with initial masses increasing from bottom to top. Solid (dotted) lines correspond to rotating (nonrotating) models.
\label{Fig:Mcc}}
\end{figure}

The masses of the convective core during the core H-burning phase for rotating and nonrotating models of different initial masses at Z=0.0004 are presented in Fig.~\ref{Fig:Mcc}. At the beginning, the convective cores in the rotating models are smaller than in the nonrotating ones. This comes from the effect of the centrifugal acceleration that supports part of the weight and thus makes the central temperature lower in the rotating models than in the nonrotating models. The shifts are larger in stars that are most easily deformed by the centrifugal force, that is, the most massive ones because they are less dense.

\begin{figure*}
\includegraphics[width=0.49\textwidth]{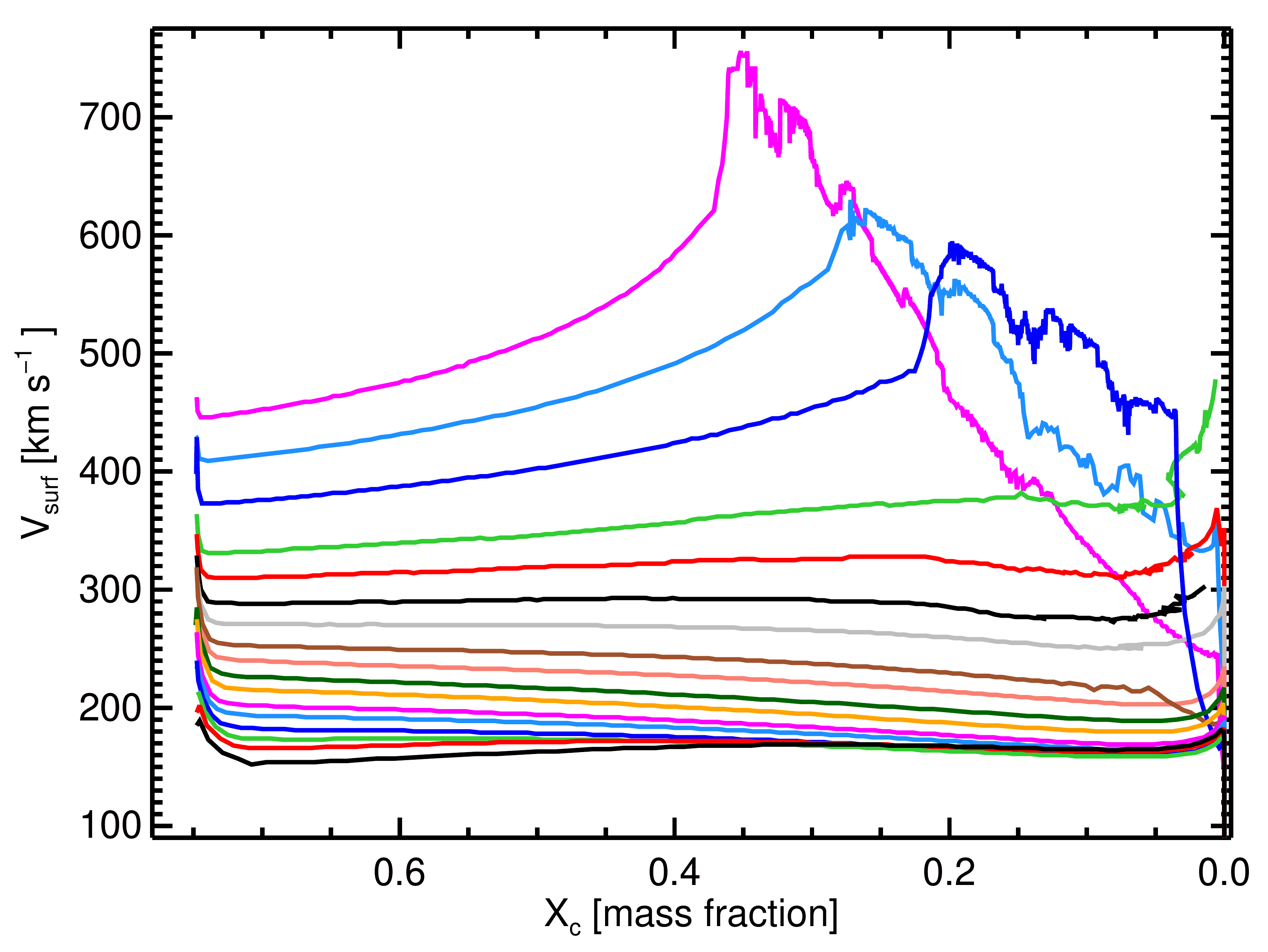}\hfill\includegraphics[width=0.49\textwidth]{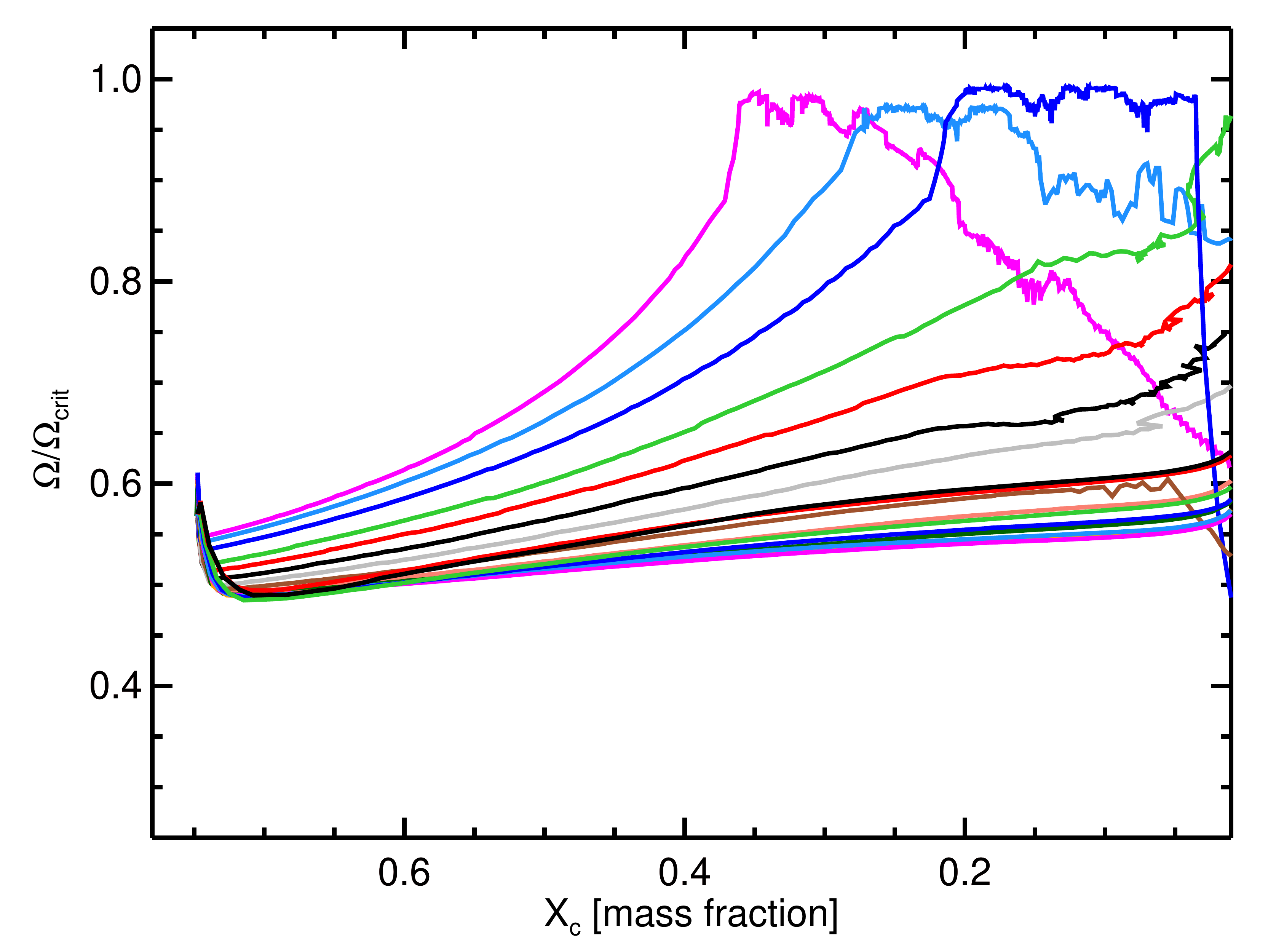}
\caption{\textsl{Left panel:} Evolution of the surface equatorial velocities during the MS phase. The mass sequence is increasing from the black bottom line ($1.7\,M_\odot$) to the upper magenta line ($120\,M_\odot$). \textsl{Right panel:} Evolution of the ratio $\Omega/\Omega_{\rm crit}$ for the same models.}
\label{Fig:Vsurf}
\end{figure*}

In general, the convective cores decrease smoothly in mass during the whole MS phase. There are however two exceptions: first in the case of the rotating 1.7 $M_\odot$ model at the very beginning of the core H-burning phase and, second, for the stars more massive than 20 $M_\odot$ at the end of the core H-burning phase. We discuss these two cases below.

% findings 
At Z=0.0004 in the nonrotating 1.7 M$_\odot$ model, the convective core decreases in mass as evolution proceeds during the MS phase (see the bottom black dotted curve in Fig.~\ref{Fig:Mcc}). When rotation is accounted for, a very different evolution is obtained (see the bottom black continuous curve in Fig.~\ref{Fig:Mcc}). First, the convective core increases in mass and then suddenly, at an age equal to about 120 Myr, drops to a value around 0.15-0.16 $M_\odot$ and remains around that value until the end of the MS phase. This behavior in the rotating model results from the following phenomena: 
\begin{enumerate}
\item The size of the convective core depends on the ratio between the nuclear energy produced by the proton-proton (pp) chains and the CNO cycle. When the
CNO cycle dominates, larger convective cores appear. This comes from the well-known fact that the energy production rate for the CNO cycle
shows a much stronger dependence on temperature ($\propto T^{17}$) than the pp chains ($\propto T^4$).  
\item The typical temperatures at the center of the 1.7 $M_\odot$ model are in a range (around $22\times10^6~\K$) where
both the pp chains and the CNO cycle contribute significantly to the nuclear energy production. Small variations of the central temperature therefore change the size of the convective core
significantly.
\item Rotation induces two counteracting effects, one favoring the importance of the CNO cycle and the other one
disfavoring it. On one side, rotational mixing allows carbon to diffuse from the envelope to the core fueling the CN cycle (we note that for this model, the ON cycle is negligible) thus favoring the CN cycle leading to large convective cores. On the other hand, rotation, by sustaining in part the weight of the stellar material through the centrifugal acceleration, tends to lower the central temperature, disfavoring the CN cycle and producing smaller convective cores. 
\end{enumerate}
At the beginning of the evolution, although the central temperature in the rotating model is smaller than in the nonrotating one, still the CN cycle dominates. However, after the stage when carbon has been depleted around the core, the timescale for fuelling carbon into the core becomes too long for allowing the CN cycle to be maintained. At this point,  the lower central temperature in the rotating model causes the energy to be produced mainly by the pp chains, and the size of the convective core drops. A similar effect was found in \citet{Georgy2013b}.

We may wonder to what extent this feature would have been different if pre-MS evolution had been computed for this model. Indeed, during the pre-MS phase some CN burning occurs in the core, modifying the abundances of these elements in the central regions of the star. However, this would occur when the star is nearly on the ZAMS and thus accounting for this phase would have a very limited impact.

The increase of the core mass at the end of the MS phase for rotating stars with initial masses above about 20 $M_\odot$  results from the action of the effective diffusion coefficient, $D_{\rm eff}$, that dominates the mixing just above the convective core (see Fig.~\ref{D7}). This coefficient attempts to capture the physics of the interaction between meridional currents and the strong horizontal turbulence responsible for the shellular rotation law \citep{CZ1992}. The coefficient is higher for massive stars than for low-mass stars. For this reason, the more massive models have more mixing, which counteracts the usual decline of the mass of the convective core, and may even stop the decline temporarily (Fig.~\ref{Fig:Mcc}). We return to this point in Sect.~\ref{Subsec:metaltime}, where we discuss how metallicity affects the internal structure of stars.

The effects of rotation on the duration of the evolutionary phases can be seen in Table~\ref{Tab:properties}. As found in \citetalias{Ekstrom2012a}, \citetalias{georgy12a} and \citetalias{Georgy2013a}, our models also show that an initial value of $\vrot/\vcrit=0.4$ increases stellar lifetimes by about 20\% compared to nonrotating models.

%=================================================================================
\subsection{Rotational velocities\label{Subsec:rotation}}

\begin{figure*}
\includegraphics[width=0.48\textwidth]{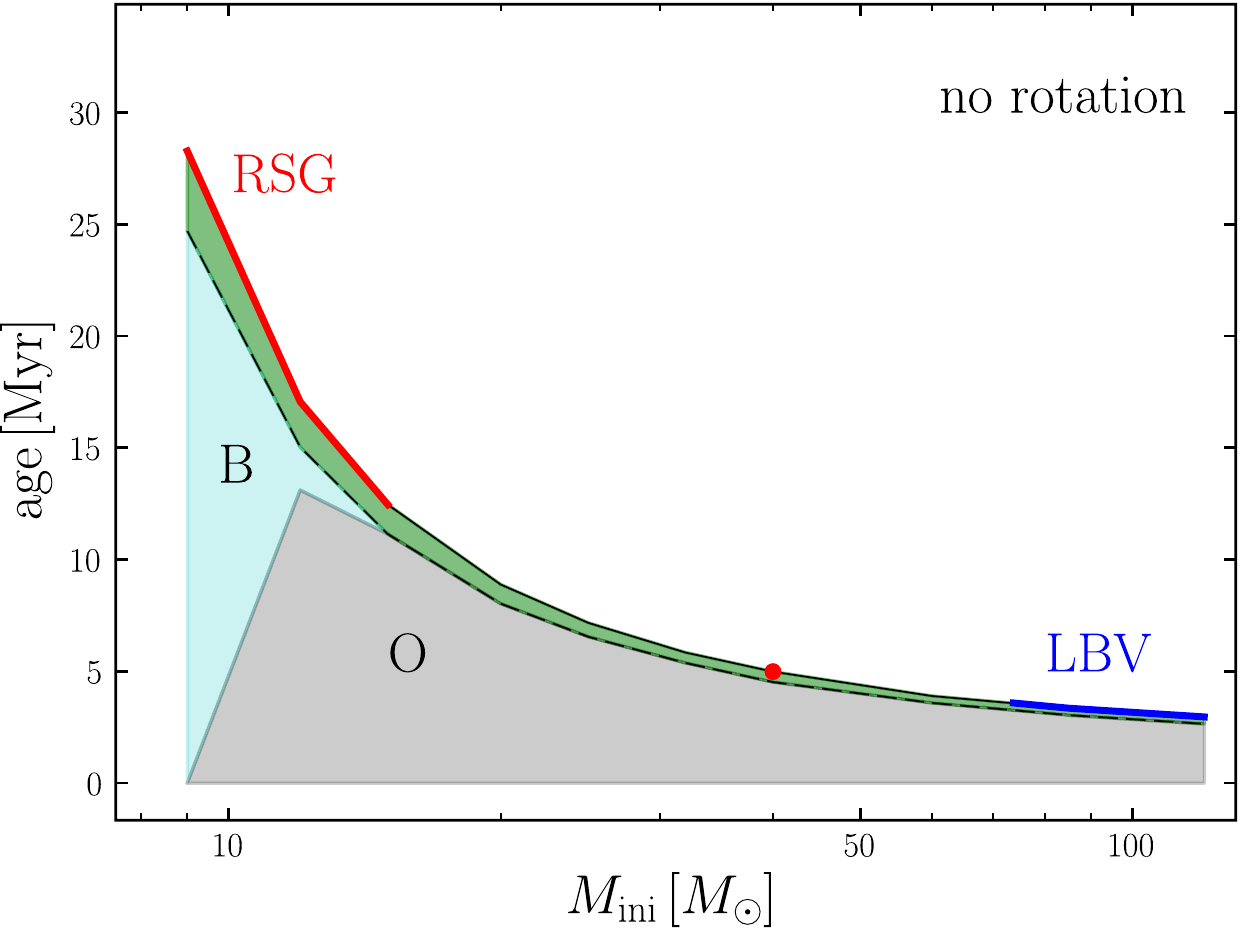}\includegraphics[width=0.48\textwidth]{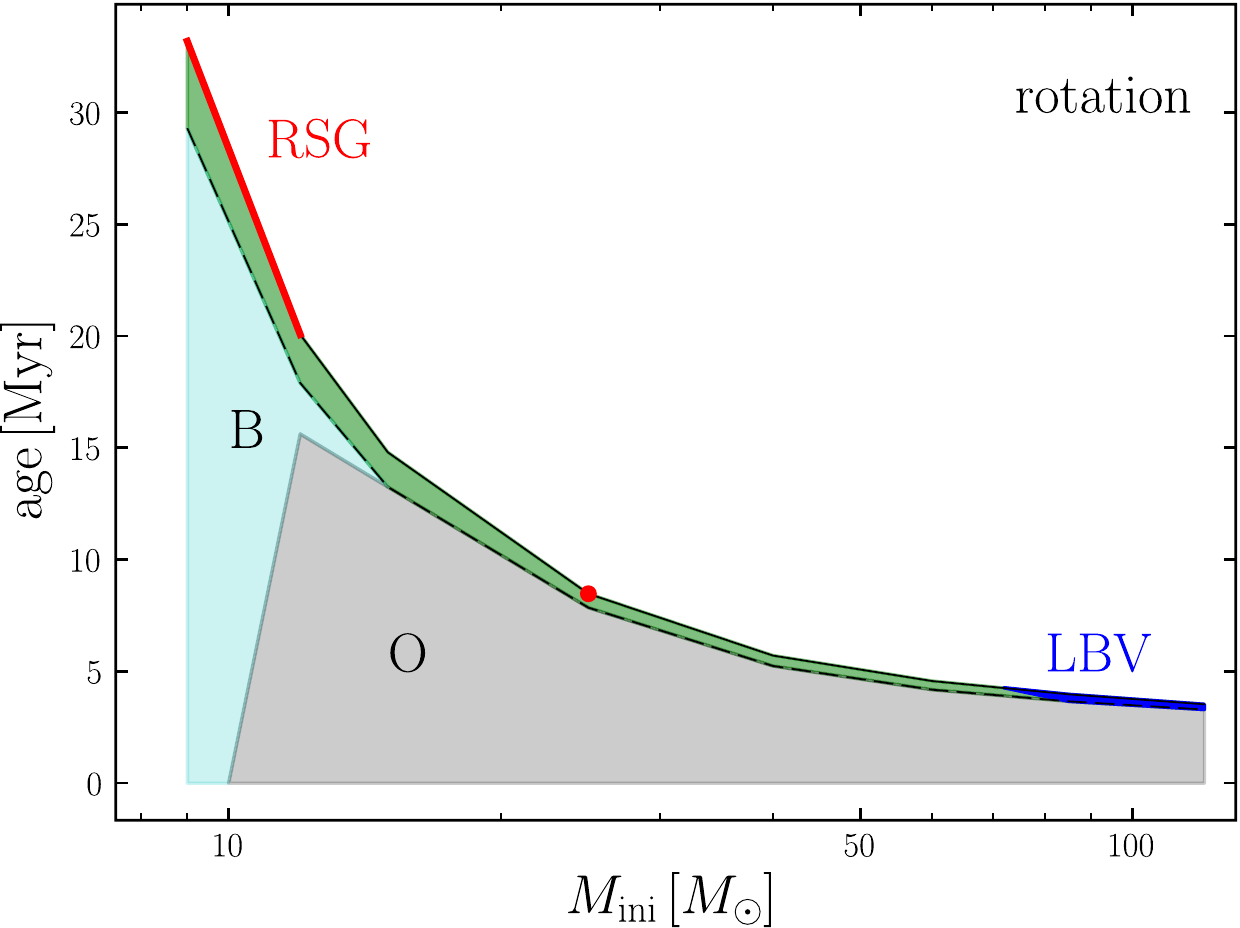}
\caption{Lifetimes during different evolutionary phases as a function of  initial mass for nonrotating (left) and rotating models (right). The upper curve corresponds to the total lifetime and the dashed black curve to the MS lifetime, which is further divided in O-type and B-type stars.The blue regions beyond the MS phase correspond to yellow-blue supergiants, red to red supergiants, and blue to LBV stars.}
\label{Fig:wr}
\end{figure*}

The initial rotational velocities of our models and their time-averaged value during the MS phase are shown in Table 1. The increase of initial rotational velocity with mass is a result of adopting a fixed ratio for the initial \vrot/\vcrit\ for all masses, since \vcrit\ increases with initial mass. 

The time-averaged value of \vrot\ depends on three competing mechanisms: expansion of the envelope, meridional currents, and mass loss. In general, the time-averaged surface rotational velocity during the MS phase ( $\bar{V}_\mathrm{MS}$) is lower than the initial surface rotational velocity. This occurs because the envelope expands during the MS, which decreases the surface velocity. The angular momentum transported by meridional currents is insufficient to counteract this effect. However, the more massive models ($\mini >  85~\msun$) show a higher value for the time-averaged  \vrot\ than the initial value of \vrot. In these models, the meridional currents that transport angular momentum from the inner to the outer regions are strong, and thus more than compensate for the decrease of the surface rotational velocity due to the envelope expansion. In general, mass loss increases during the MS for the massive models because of the increase in luminosity. Still, the absolute value of the mass-loss rate during the MS phase is very modest at these low metallicities, and thus removes only a very small amount of angular momentum from the surface. 

Figure \ref{Fig:Vsurf} presents the evolution of surface rotational velocities during the MS phase. For all the masses below about 20 $M_\odot$, the surface velocity remains constant during the MS phase. If there was local conservation of the angular momentum,  the surface velocity should have decreased during that phase since the envelope expanded during the MS. The models indicate that angular momentum is transported from the core to the surface, compensating for the envelope expansion. 

For the most massive stars, the surface velocity increases during the MS phase. This occurs because high-mass stars have less dense radiative envelopes and therefore stronger meridional currents that can transport angular momentum from the interior of the star to the surface. At a given point, the star reaches a surface velocity that is high enough for the centrifugal acceleration to compensate the gravity at the equator \citep{mm_omega00}. In this case, we assume that the layers that are overcritical are removed, as in Paper \citetalias{georgy12a}. The three most massive models reach the critical limit during the second half of the MS and experience these mechanical mass losses. The stars are still hot O-type stars, and would be observationally associated with Oe stars if the disk became massive enough. In this case, emission-line components should be seen in the H lines.

An interesting consequence from our models is that the value of the surface velocity during the MS phase for stars with masses below 20 $M_\odot$ would give a reasonable estimate of their initial rotation. We note that the current models assume that the star initiates its evolution on the ZAMS rotating as a solid body. As was already discussed in previous studies \citep[e.g.,][]{Haemmerle2013},  a rapid redistribution of the angular momentum through meridional currents occurs at the very beginning of the calculations, transporting angular momentum from the surface to the central region. As a result, the surface velocity shows a rapid drop at the beginning of the evolution. Because the initial internal rotation profile has weak effects on the further evolution of the star, starting with a nonsolid body rotation profile would produce similar results for the subsequent evolution provided that the total angular momentum is kept constant \citep{Haemmerle2013}.

%=================================================================================

%=================================================================================
\subsection{Surface abundances\label{Subsec:abund}}

To illustrate the evolution of surface abundances and the effects of rotational mixing, the evolutionary tracks shown in Fig.~\ref{Fig:HRDgen}  are color-coded according to the surface abundance ratio of nitrogen to hydrogen in logarithm scale (by number and normalized to a number of hydrogen atoms equal to 10$^{12}$). The evolution of the surface abundances is mainly affected by two mechanisms, mixing (rotational or convective) and mass loss. 

The nonrotating models predict no change in the surface abundances during the MS phase for all the mass domain explored here. Below 4 $M_\odot$, the surface nitrogen-to-hydrogen ratio changes by at most 0.4 dex as a result of the first dredge-up. Between 4 and 40 $M_\odot$, there are no changes in the surface abundance during core He-burning. Stars with masses between 9 and 15 $M_\odot$ become red supergiants after the end of the core He-burning phase, developing a convective envelope and showing CNO-processed products at their surfaces as a result of dredge up. Stars with masses between 20 and 40 $M_\odot$ do not change their surface composition during their lifetimes. These stars do not develop a deep convective layer, and therefore no dredge up occurs. Above 40 $M_\odot$, the ratio between N and H changes by more than 1.0 dex as a result of mass loss by stellar winds, which reveals the layers that have been enriched with CNO-processed material.

The rotating models show rapid changes in the N/H ratio during the MS phase due to rotational mixing. At any given evolutionary stage, these effects are more pronounced for higher masses due to the combined effects of mass loss and rotational mixing, which are both more effective at higher masses. We note that at the end of the 7 $M_\odot$ model, the N/H ratio increases by 0.5-0.6 dex, which causes the track to shift to slightly cooler effective temperatures. This occurs because this particular model has a deep convective envelope and also mixes C, N, and O from the core into the base of the convective envelope. More massive models do not show this effect because the convective envelope is not deep enough. Less massive models were stopped when the core became degenerate, and could still show increased CNO abundances in the envelope at later evolutionary phases.

In addition, our models show that Cepheids originating from nonrotating or slowly rotating stars are predicted to show no change of the N/H ratio at their surfaces, except in the lowest initial mass range crossing the instability strip after the first dredge-up occurred.

%=================================================================================
\begin{figure*}
\centering \includegraphics[width=0.96\columnwidth]{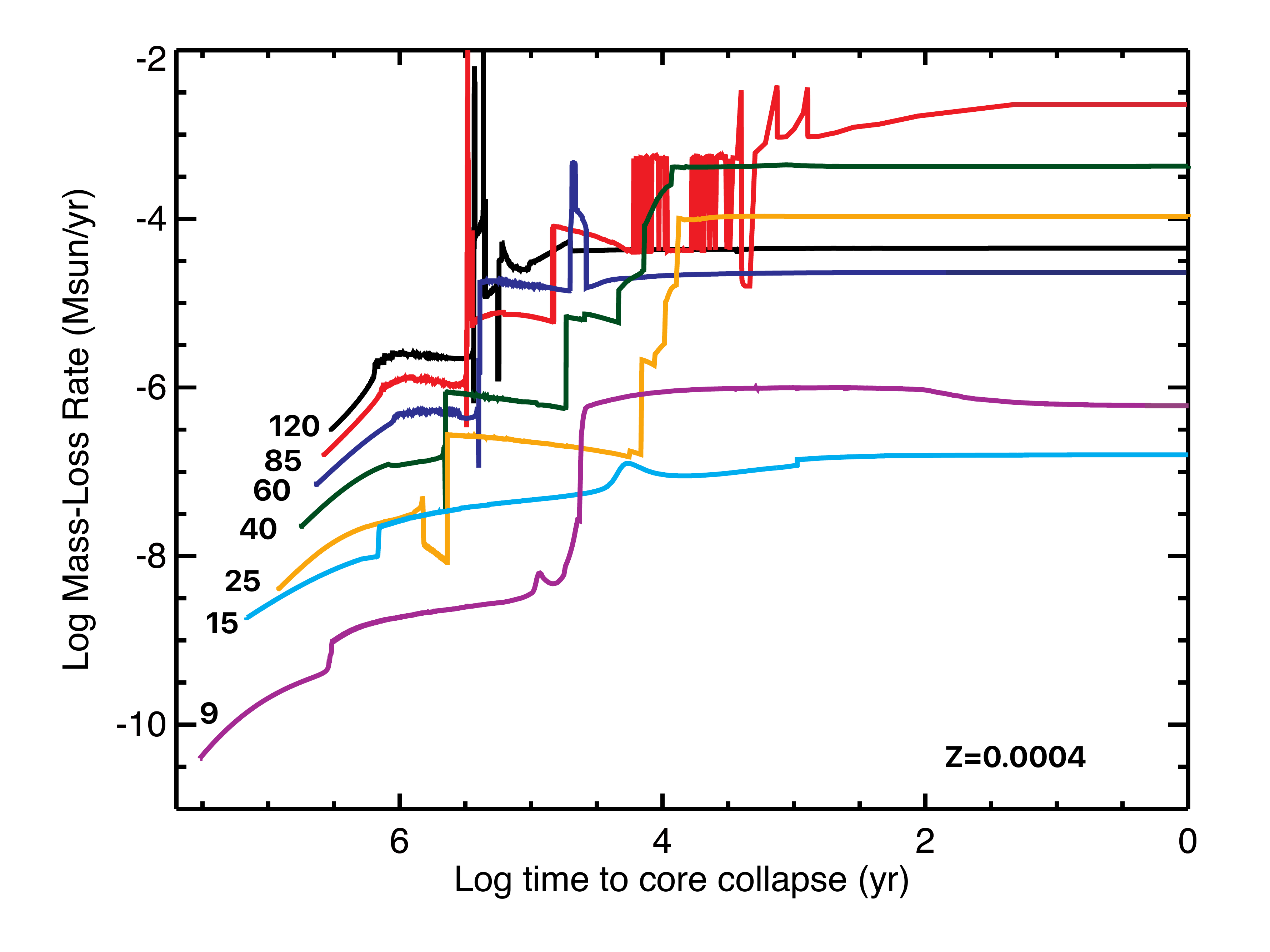}\includegraphics[width=0.96\columnwidth]{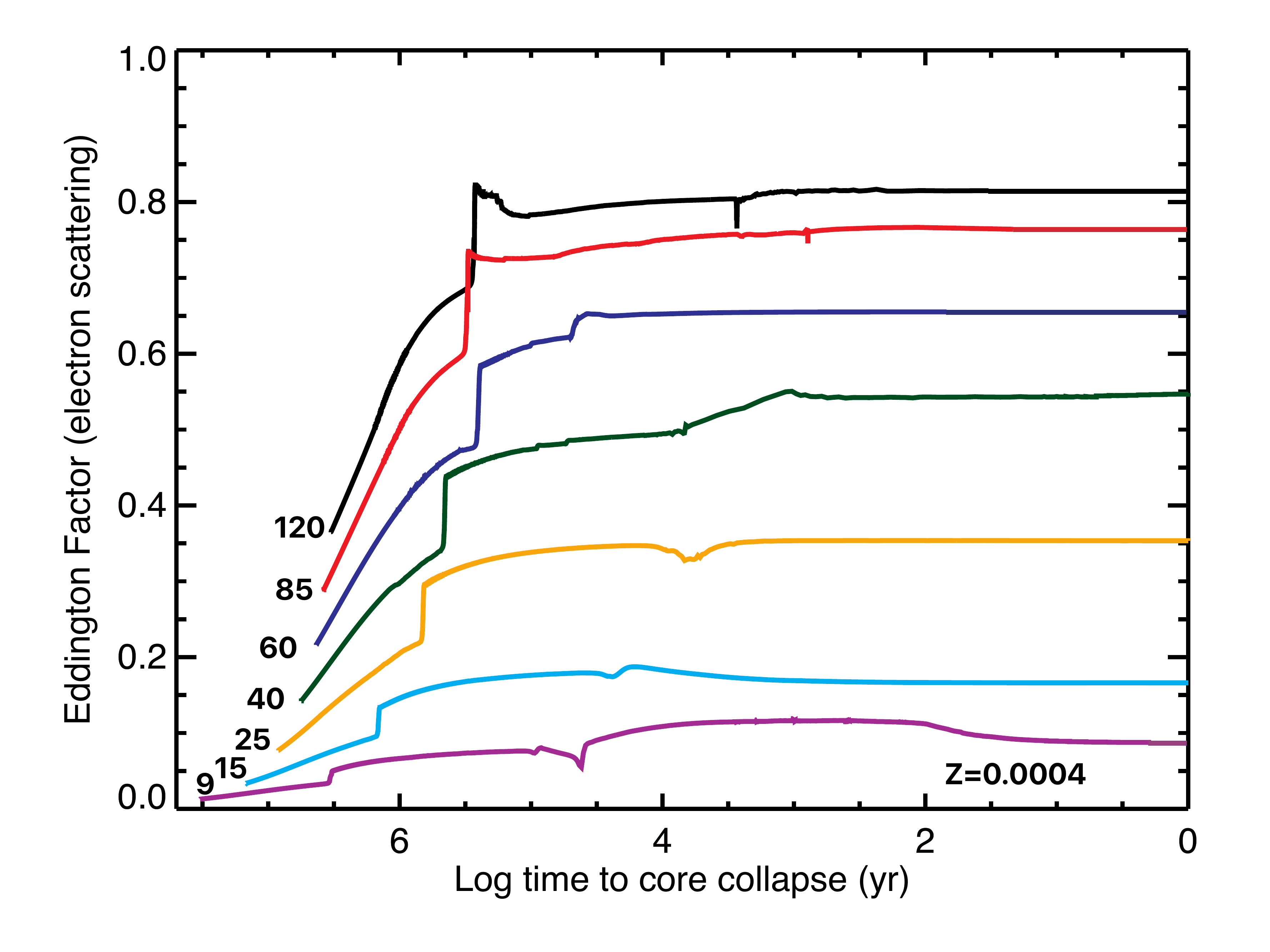}
\caption{Mass-loss rates (left panel) and Eddington parameter \eddesc\ (right panel) as a function of time to core collapse for selected rotating models (9, 15, 25, 40, 60, 85 and 120~\msun). The labels indicate the initial mass.}
\label{Fig:Mdot}
\end{figure*}

\subsection{Massive star populations from single stars at very low metallicity \label{Subsec:pop}}

Our stellar evolution models allow an estimate of the content of stellar populations in very-low-metallicity environments. For consistency with Papers I and III, here we assume the calibrations between spectral types and surface properties as in \citet{georgy12a}, and broadly define that O-type stars have $\log (\teff/\K) > 4.477$ and a hydrogen mass fraction at the surface greater than 0.4, yellow-blue supergiants (YSG-BSG)  have  $\log (\teff/\K)$ in the range 3.660 and 4.477, and RSGs have $\log (\teff/\K) < 3.660$. We also assume that the region of the stellar evolution tracks with  $\log (\teff/\K) > 4.477$ and a hydrogen mass fraction at the surface lower than 0.4 correspond to WR stars. Coupled stellar evolution and atmospheric modeling would be needed to provide a full view on the stellar properties \citep{gme14,martins17}, but we defer these computations to future work. In particular, because of the presence of a strong stellar wind towards the end of the evolution, we expect that part of the stars that we classify here as YSG-BSGs based on $\teff$ and abundances could instead show a spectrum more similar to those of LBV stars \citep{gme13,gmg13}.

Figure~\ref{Fig:wr} compares the model predictions for the lifetimes of the different spectroscopic classes as a function of initial mass. The vast majority of the stellar lifetime is spent as an OB-type star, with very short RSG and LBV lifetimes. The rotating models have longer O-type and LBV phases than nonrotating models. The RSG lifetimes are very short for both families of models, with the rotating (nonrotating) models becoming RSGs only up to 12 (15) $M_\odot$. 

 \subsection{Mass-loss history and the Eddington parameter \label{Subsec:mdot}}

Figure~\ref{Fig:Mdot} shows the evolution of the mass-loss rate and Eddington parameter for the 9, 15, 25, 40, 60, 85 and 120~\msun\ models with rotation. Here, we consider only electron scattering opacity ($k_\mathrm{es}$), with the Eddington parameter for electron scattering being
\begin{equation}
\eddesc\ =\frac{k_\mathrm{es} L }{4 \pi G c  M} ,
\end{equation}
where $L$ is the radiative luminosity, $M$ the stellar mass, and the constants have their usual meanings. Thus, the values of \eddesc\ are lower limits for the Eddington parameter since other opacity sources may be relevant. The values of \mdot\ are modest during the majority of the core H-burning phase ($\mdot \lesssim 3\times10^{-6}~\msunyr$), even for stars with initial masses of 120~\msun. The massive star models show a substantial increase in \mdot\ by about a factor of ten after the core H-burning phase, when their effective temperatures decrease and the stars cross the first bi-stability limit around $\teff=21000-25000~\K$ \citep{Vink2001}. The most massive models cross the second bistability jump ($\teff=10000-15000~\K$) towards the end of the core He-burning phase, causing \mdot\ to increase by roughly another factor of ten. Because the models above 40~\msun\ have $\log (\teff/\K) > 3.900$, the code switches from the \citet{Vink2001} to the \citet{dejager88} recipe, which produces a short period of increased mass loss.

In the last 10,000 years before core collapse, the \mdot\ values range from $10^{-7}~\msunyr$ for the 15~\msun\ model to a few times $10^{-2}~\msunyr$ for the 85~\msun\ model. Based on the mass-loss recipes adopted here, stars at very low metallicity such as $Z=0.0004$ could show substantial mass loss before core collapse, which could affect both the chemical yields, explosion properties, and supernova appearance \citep{yusof13,hirschi17}.

Our models indicate that massive stars can approach the Eddington limit at $Z=0.0004$ (Fig.~\ref{Fig:Mdot}). Their values of \eddesc\ range from 0.08 to 0.37 at the ZAMS for 9 and 120~\msun\ models, respectively. As the star evolves, \eddesc\ increases substantially for $\mini \ge 25~\msun$, ranging from 0.3 to 0.8 for the 25 and 120~\msun\ models, respectively, during the post core H-burning phase. These values remain high until core-collapse (Fig.~\ref{Fig:Mdot}), which may cause eruptive mass loss in the last $10^{4}$ yr of evolution and produce bright interacting supernovae.

\subsection{Properties at advanced stages of the evolution \label{Subsec:mass}}

\begin{figure}
\center
\includegraphics[width=0.49\textwidth]{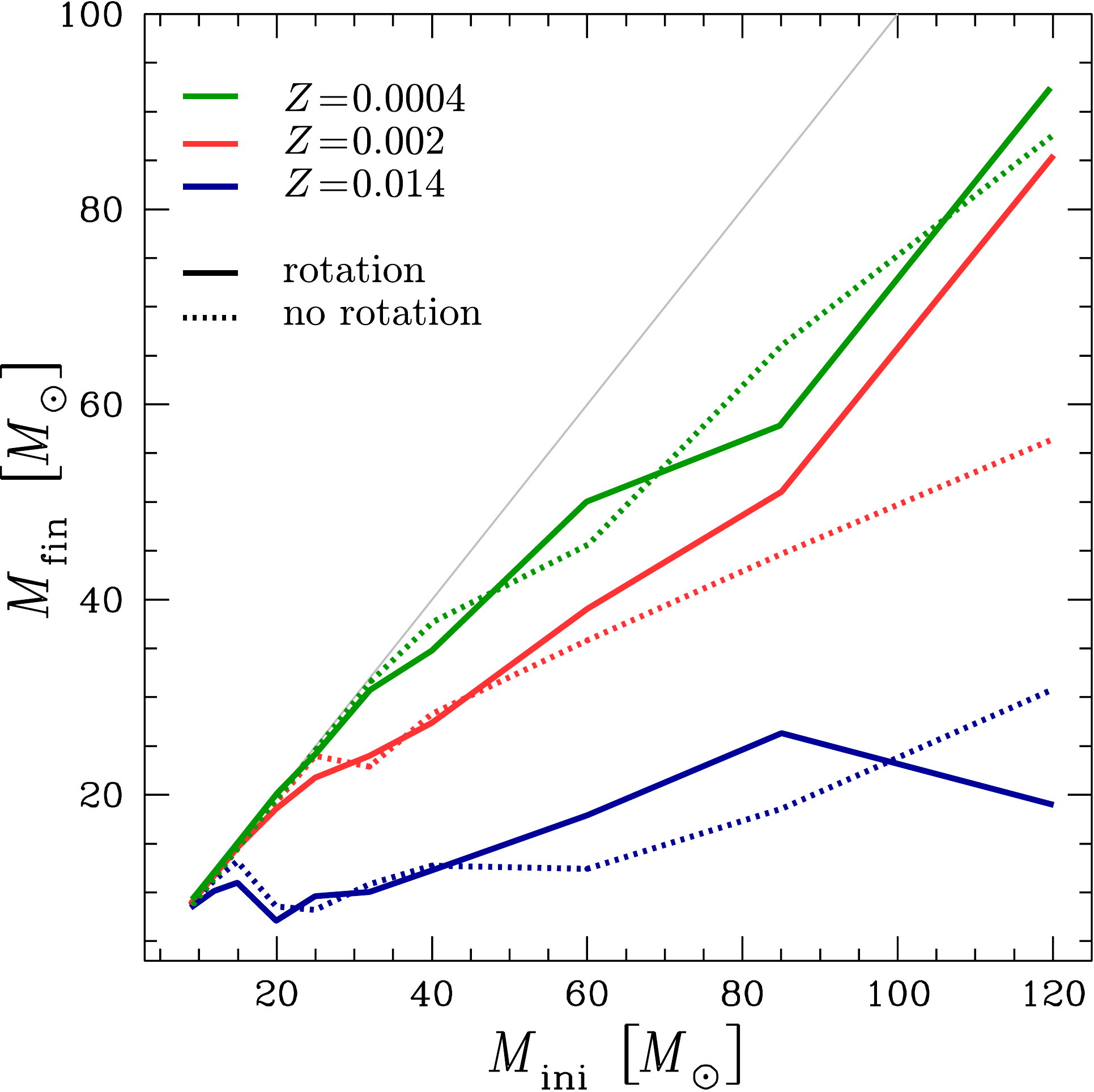}
\caption{Final mass ($M_\mathrm{fin}$) of the models as a function of the initial mass (\mini) at various metallicities, with and without rotation. The 1:1 relationship is indicated as a thin gray line.}
\label{Fig:lifetime}
\end{figure}

In this section we investigate the final mass, core mass, abundance structure, angular velocity structure, and surface properties at advanced stages of the evolution. Figure~\ref{Fig:lifetime} shows the final masses of our rotating and nonrotating models. At $Z=0.0004$, their initial-mass-to-final-mass ratios are close to 1:1, shown by the diagonal line shown in Fig.~\ref{Fig:lifetime}. This is caused by the weakness of the stellar winds at low metallicity. Figure~\ref{m120} presents the chemical abundance, radius, and angular velocity profiles as a function of the Lagrangian mass coordinate for rotating models with initial masses of 120, 20, 7, and 2.5 $M_\odot$. They are representative of the different initial mass regimes, and we discuss selected properties below.

\subsubsection{The 120 $M_\odot$ model}
\begin{figure*}
\centering
\includegraphics[width=0.43\textwidth]{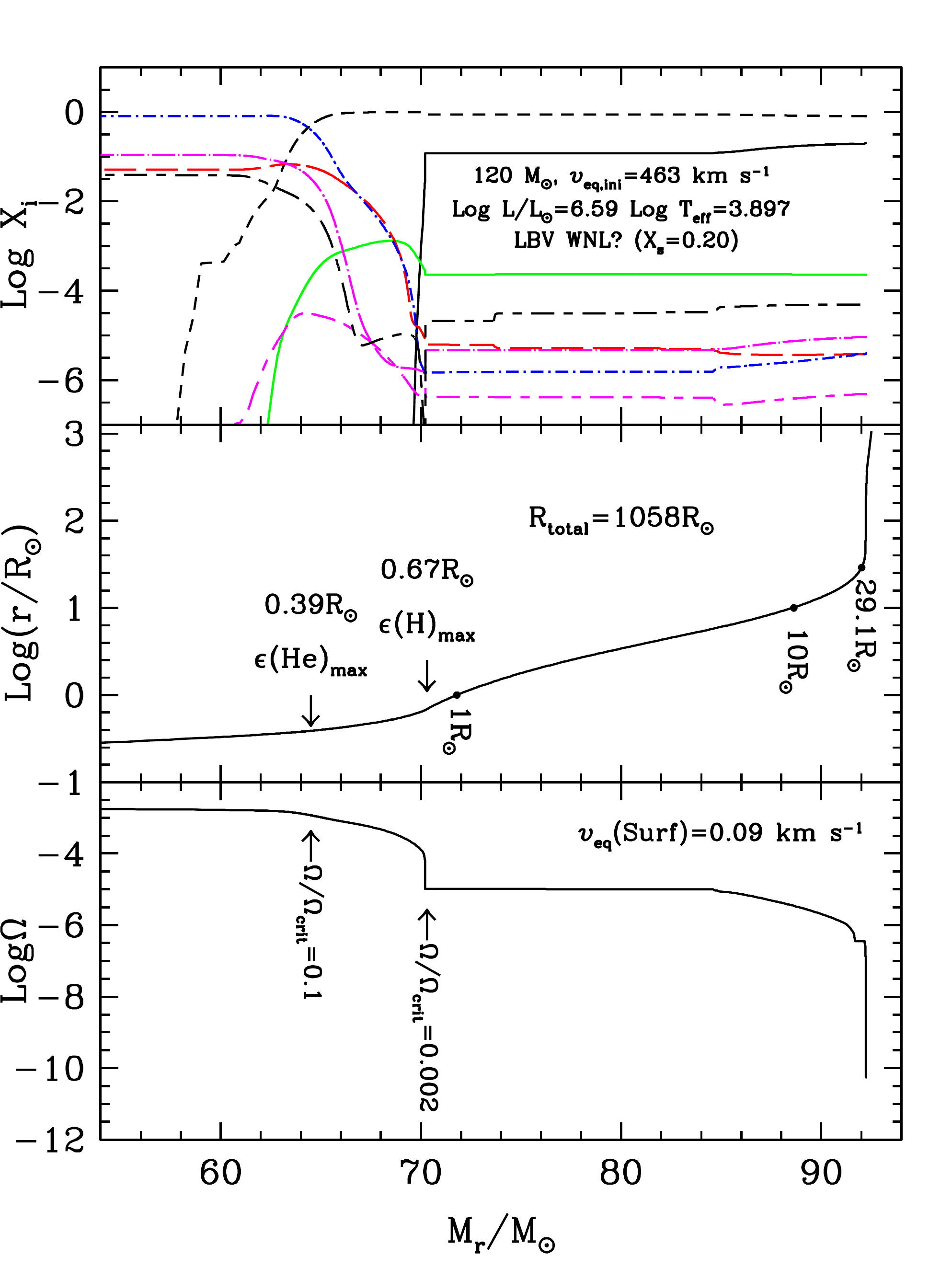}\includegraphics[width=0.43\textwidth]{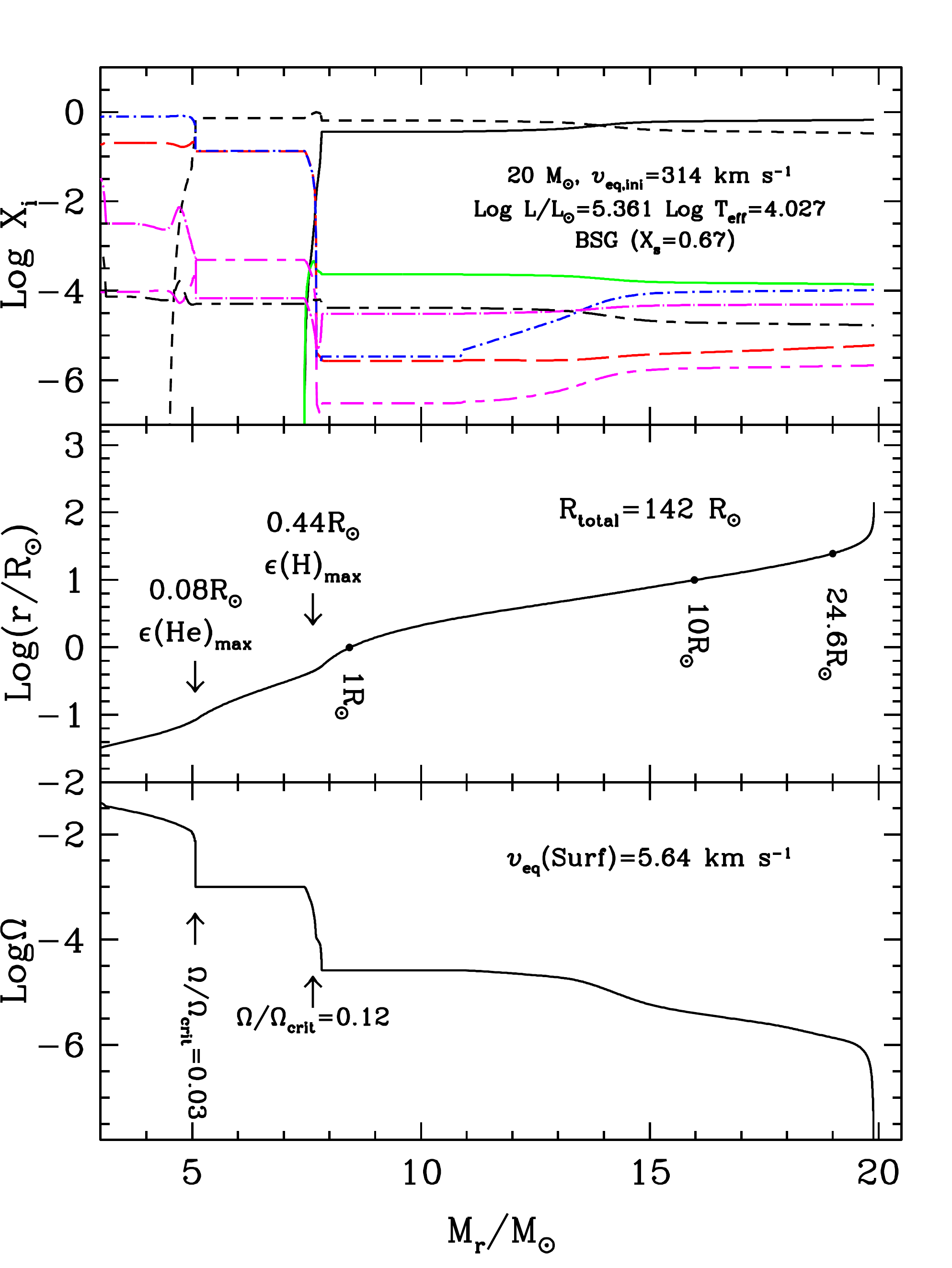}
\includegraphics[width=0.43\textwidth]{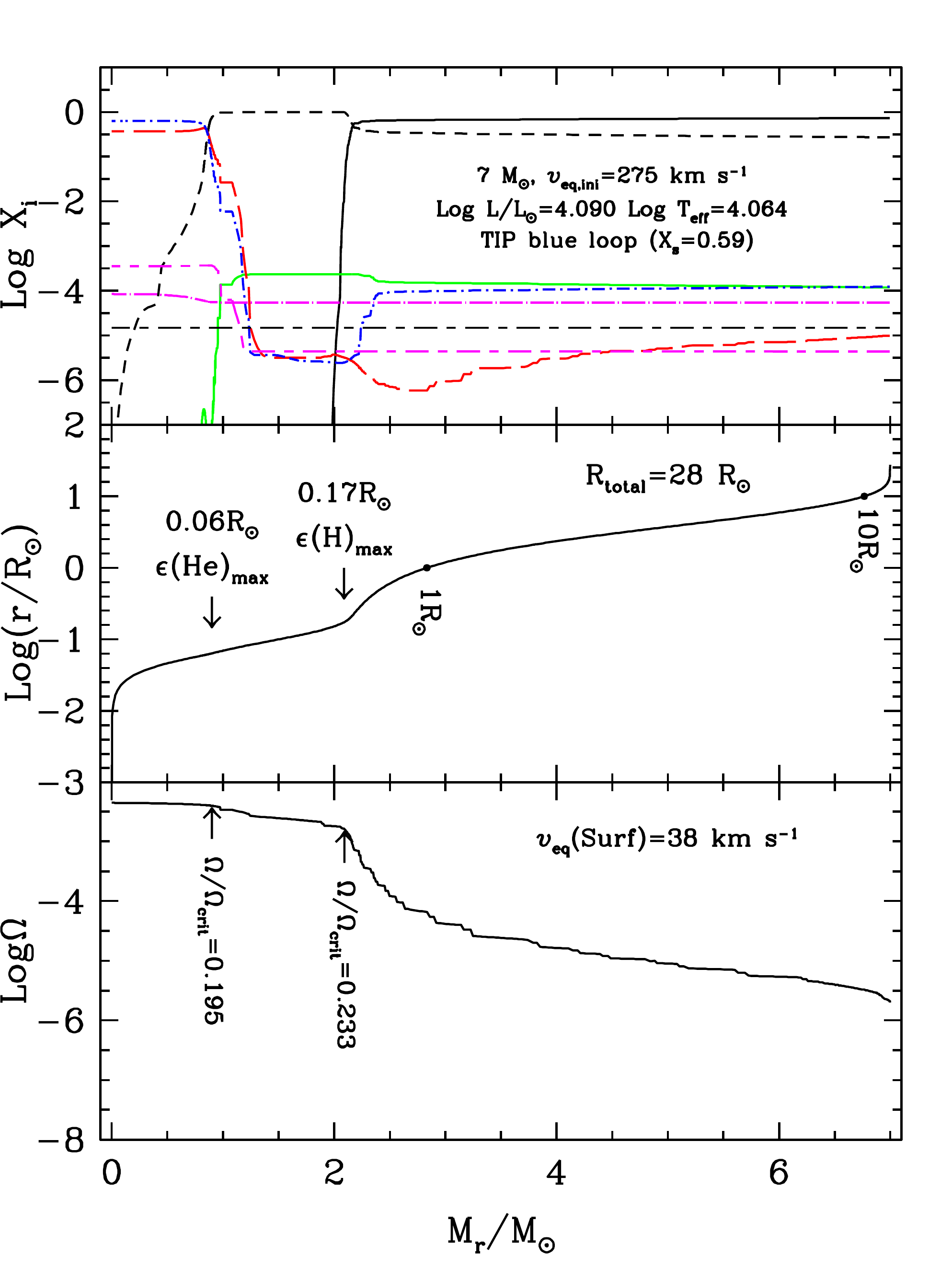}\includegraphics[width=0.43\textwidth]{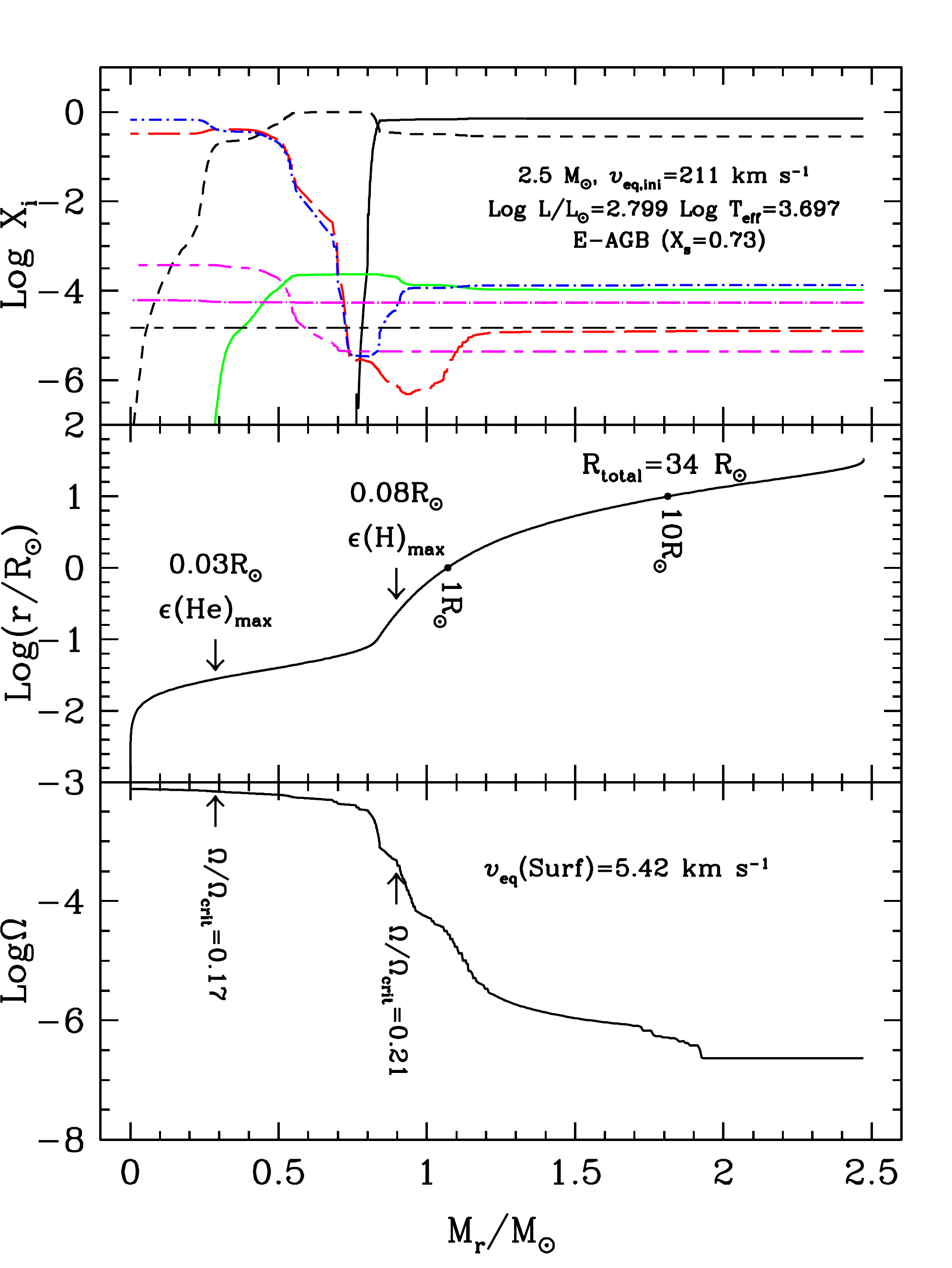}
\caption{Structure of rotating models in advanced phases of the evolution for initial masses 120~\msun\ (top left; end of C-burning phase), 20~\msun\ (top right; end of C-burning phase), 7~\msun\ (bottom left; early AGB phase), and 2.5~\msun\ (bottom right; early AGB phase). We show three panels for each initial mass. The top panel shows the chemical structure as a function of the Lagrangian mass coordinate $M(r)$, with the mass fractions of H (black solid line), He (black dashed), $^{12}$C (red long dashed), $^{14}$N (green solid), $^{16}$O (blue dot-dashed), $^{20}$Ne (magenta dot-long dashed), $^{22}$Ne (magenta short-long dashed), and $^{24}$Mg (black short-long dashed). The middle panel shows the radius as a function of the Lagrangian mass coordinate. We also indicate the location of the maximum energy production of the He- and H-burning shells. The lower panel, the variation of the angular velocity as a function of the Lagrangian mass coordinate, with the labels indicating the ratio of the angular velocity to the critical velocity at selected positions.}
\label{m120}
\end{figure*}

The 120 $M_\odot$ clearly shows a production of primary nitrogen. This appears in the layers containing the products of H-burning via the CNO-cycle (mass coordinates 64--70 $M_\odot$), where the nitrogen mass fraction reaches values up to 0.001. This is four times the sum of the initial mass fractions of CNO elements ($\sim$0.00025) indicating that rotation induces the production of primary nitrogen. The physics of this process is  discussed in \citet{meynet02}. At low Z, there is an efficient transport of $^{12}$C from the He-burning core to the H-burning shell, where the CNO cycle converts $^{12}$C  to $^{14}$N. The other models shown in Fig. \ref{m120} present no significant primary N production, since no C is transported efficiently from the core to the H-burning shell.

Before core collapse, the 120 $M_\odot$ star has its mass relatively concentrated in the central regions. From the middle panel we can see that nearly 99.5\% of the total mass is enclosed in 30 R$_\odot$. The remaining 0.5\% of the mass extents over more than 1000 R$_\odot$. The H mass fraction at the surface is 0.2, but the relatively low effective temperature indicates that the star would be spectroscopically similar to an AF-type supergiant or LBV. The final fate of this star depends strongly on mass loss, which is extremely uncertain in this parameter range. We caution the reader that no theoretical or empirical mass-loss recipe exists for yellow hypergiants and LBVs at these high luminosities (see \citealt{smith14araa} for a review on massive star mass loss). In our models, we extrapolate the values predicted by the \citet{dejager88} recipe to high luminosities, and so the final mass and evolution should be interpreted with this in mind. 

At the end of C-core burning, the helium core has a mass of about 70 $M_\odot$. This is in the range of values between 64 and 133 $M_\odot$ where, according to \citet{HW2002}, a pair-instability supernova (PISN) would occur. During such events, the star is completely destroyed and its matter dispersed. If the AF-supergiant/LBV progenitor ejected significant amounts of material, the SN ejecta would crash into this circumstellar medium and produce an interacting SN. Depending on the CSM mass and SN ejecta velocity and mass, the star could produce a type II superluminous SN.

The 120~\msun\ model shows a very strong contrast in the angular rotation profile, with the core rotating faster than the surface by  nearly seven orders of magnitude (Fig.~\ref{m120}c). This is caused by angular momentum loss from surface layers, which is a result of the substantial mass loss. Still, the central regions are rotating slowly, at about 10\% of the critical velocity at the end of C burning. Before the explosion, these layers will collapse and $\Omega/\Omega_{\rm crit}$ will increase as $1/\sqrt{r}$ since the radius decreases. The radius would have to decrease by a factor of 100 in order for the inner layers to achieve the substantial value of  $\Omega/\Omega_{\rm crit}$ and possibly affect the properties of the explosion. This is unlikely to happen in the 120~\msun\ model, as the star would probably explode due to pair instability before near critical rotation is reached. Also, the models do not include the effects of magnetic fields, and our results are upper limits concerning the magnitude of the core rotation.

\subsubsection{The 20 $M_\odot$ model}

The rotating 20 $M_\odot$ model ends its lifetime as a blue supergiant. We estimated the remnant mass based on the CO core mass following \citet{Georgy2012}. We find that a neutron star is expected to be formed, probably accompanied by a SN II given the mass of the H envelope. The envelope is not substantially extended, favoring the appearance of a SN IIL. 

We obtain that the core of this star contains a high amount of angular momentum. Assuming that all core angular momentum would remain locked in the neutron star, we find that the neutron star would spin with a period of $6\times10^{-6}$~s. This value is about two orders of magnitude higher than the critical rotational period of a neutron star (between 4.4 and $6.50\times10^{-8}$~s; see \citealt{Georgy2012} for more details on those estimates). At face value, this discrepancy would imply that either our models are missing a braking mechanism that would decrease the angular momentum of the core and make it compatible with a rapidly spinning neutron star, or that significant angular momentum loss occurs during the explosion and proto-neutron star formation, or both. In any case, the high angular momentum of the core could potentially have a significant impact on the core collapse and explosion properties. Similarly to the case of the 120~\msun\ model, we caution that the estimates of core rotation are upper limits since we do not include magnetic fields.

\subsubsection{The 7 $M_\odot$ model}

The rotating 7 $M_\odot$ model shown in Fig.~\ref{m120} has just reached the end of the core He-burning phase and is still in the blue. The core has an angular velocity that
is nearly three orders of magnitude larger than that of the surface. If the radius of the region below the helium-burning shell contracts to a typical white dwarf radius, ($\sim 0.01~\rsun$,) this would imply
that the surface rotation of the white dwarf would be around half the critical angular velocity\footnote{$(\Omega_{\rm f}/\Omega_{\rm crit,f})/(\Omega_{\rm i}/\Omega_{\rm crit,i}) \propto \sqrt{r_{\rm i}/r_{\rm f}}$, where the subscript $i$ is for initial, {\it i.e.,} before the contraction, and $f$ is for final, after the contraction.}.

\subsubsection{The 2.5 $M_\odot$ model}

We computed the evolution of the rotating 2.5 $M_\odot$ star until the early AGB phase. The contrast between the angular velocity of the core and that of the envelope is more than 3 dex. In further evolutionary stages, the contraction of the core below the He-burning shell from 0.03 to 0.01 R$_\odot$ would make the ratio $\Omega/\Omega_{\rm crit}$ increase from 0.17 to 0.29. In contrast with the more massive stars, the core of the 2.5~\msun\ star should not produce a compact remnant with a substantial amount of angular momentum.  This is caused by two effects. First, the low-mass stars begin their evolution with a lower amount of angular momentum than the more massive stars, since we assumed a constant ratio of $\vrot/\vcrit$ for the full mass range discussed in this paper. Second, the processes that transport angular momentum from the core of the envelope act on much longer timescales in low-mass stars than in high-mass stars, which more than compensates for the lower efficiency of angular momentum transport in low-mass versus high-mass stars.

\section{Effects of metallicity on stellar evolution \label{Sec:metal}}

This paper extends the database of updated Geneva stellar evolution models down to a metallicity of 1/35 solar. Since the current models were computed with the same physical assumptions as for the Z=0.014 (Paper I) and Z=0.002 grids (Paper III), they form a homogeneous dataset to investigate the effects of metallicity on stellar evolution. In this section, we discuss the implications of our models in understanding how the HR diagram, rotational velocities, surface abundances, and advanced evolutionary stages are affected by metallicity. 

\subsection{Evolution in the HR diagram}

\begin{figure}
\includegraphics[width=0.99\columnwidth]{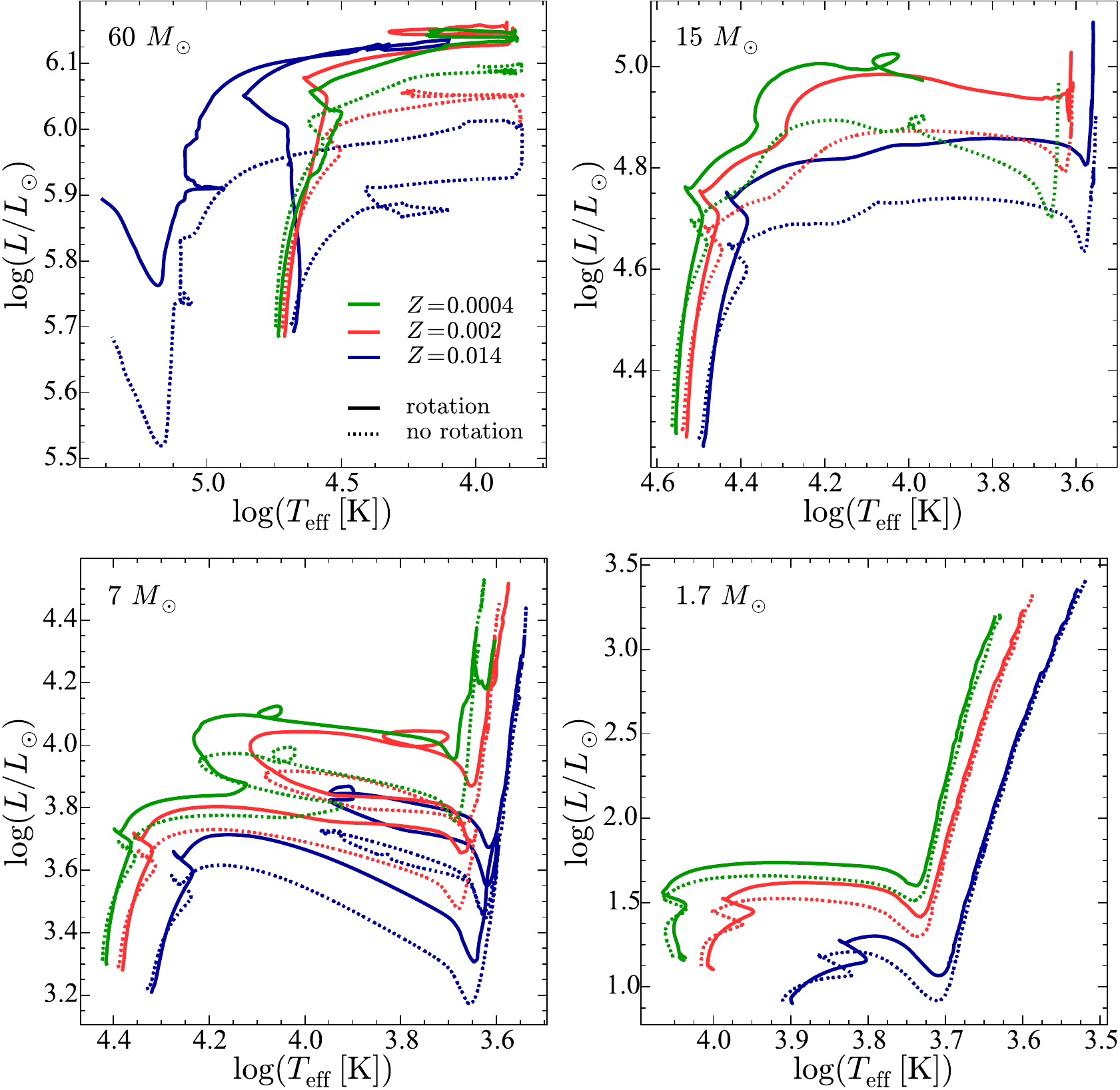}
\caption{Evolutionary tracks in the HR diagram for the $60$, $15$, $7$, and $1.7\,M_\odot$ models at various metallicities.}
\label{Fig:HRDZ}
\end{figure}

Our models show that metallicity strongly affects the stellar tracks in the HR diagram, confirming previous results \citep[e.g.,][]{schaller92,meynet94,bono2000,chieffi01,Demarque2004, Pietrin2004, Bressan2012, brott11, chieffi13, Choi2016,yoon17,eldridge17}. Figure~\ref{Fig:HRDZ} shows the metallicity effects on stellar evolutionary tracks in the HR diagram for representative models (1.7, 7, 15, and 60~\msun). As is well known, decreasing the metallicity shifts the ZAMS bluewards, and the shift is stronger for stars with lower initial mass. This occurs because stars of high initial mass are more dominated by electron-scattering opacity, while stars with low initial mass are more dominated by bound-bound and bound-free opacities \citep[see e.g.,][]{Mowlavi1998}. The metallicity dependence of the bound-bound and bound-free opacities is much steeper than the electron-scattering opacities, explaining why low-mass stars are relatively much hotter and more luminous at low metallicity compared to high-mass stars.

More generally, this metallicity effect is seen not only in the ZAMS but throughout the evolution, and is more pronounced in cool regions of the HR diagram in which the opacity is highly dependent on metallicity.  This is the case for evolved phases when the star is a red giant,  asymptotic giant branch, or red supergiant, as can be seen in the 1.7, 7 and 15 $M_\odot$ models (Figs.~\ref{Fig:HRDZ}b--~\ref{Fig:HRDZ}d). However, the  late evolution of the 60~\msun\ model is predominantly governed by mass loss and not by changes in opacity as a function of metallicity. 

The metallicity effects on blue loops are apparent in the 7~\msun\ model (Fig.~\ref{Fig:HRDZ}c), which shows an extended blue loop at solar metallicity. The Z=0.0004 model shows a more modest blue loop, starting at a much hotter effective temperature ($\log [\teff/\K] = 4.1$) than the solar metallicity model ($\log [\teff/\K] = 3.6$). This happens because the core of the Z=0.0004 model is more compact and has a higher central temperature than the Z=0.014 model, implying that a smaller contraction is needed to reach the temperature for activating the 3$\alpha$ reactions. Thus, less energy is released by the core contraction at the end of core H burning, producing a less pronounced mirror effect in the envelope and causing a higher \teff. 

\begin{figure}
\includegraphics[width=0.95\columnwidth]{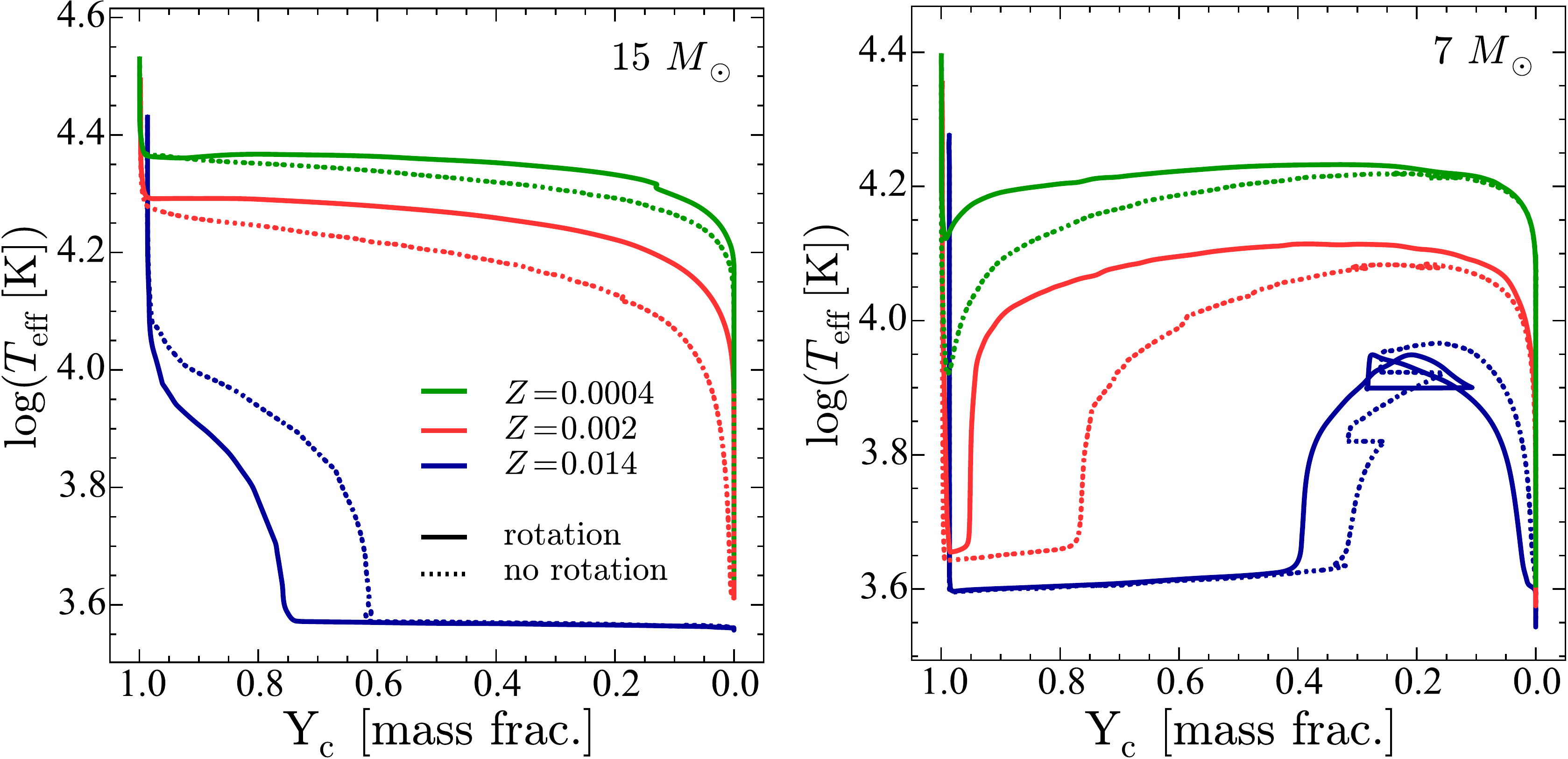}
\caption{$T_{\rm eff}$ as a function of the central He mass fraction for the $15$ and $7\,M_\odot$ at various metallicities.}
\label{Fig:YTeffZ}
\end{figure}

Figure \ref{Fig:YTeffZ} illustrates this point in another way by plotting $T_{\rm eff}$ as a function of the central He mass fraction for the $15$ and $7\,M_\odot$ models at various metallicities. While at Z=0.0004 nearly the entire core He-burning phase occurs at high effective temperatures ($\teff > 8000~\K$), only a small fraction of this phase has $\teff > 8000~\K$ at Z=0.014.

As previously noted,  a consequence of this effect is that Z=0.0004 Cepheids with masses above 4-5 $M_\odot$ only form during the first crossing of the HR gap. Thus, secular evolution produces Cepheids in that mass range for which the pulsation period only increases with age.

\begin{figure}
\includegraphics[width=0.95\columnwidth]{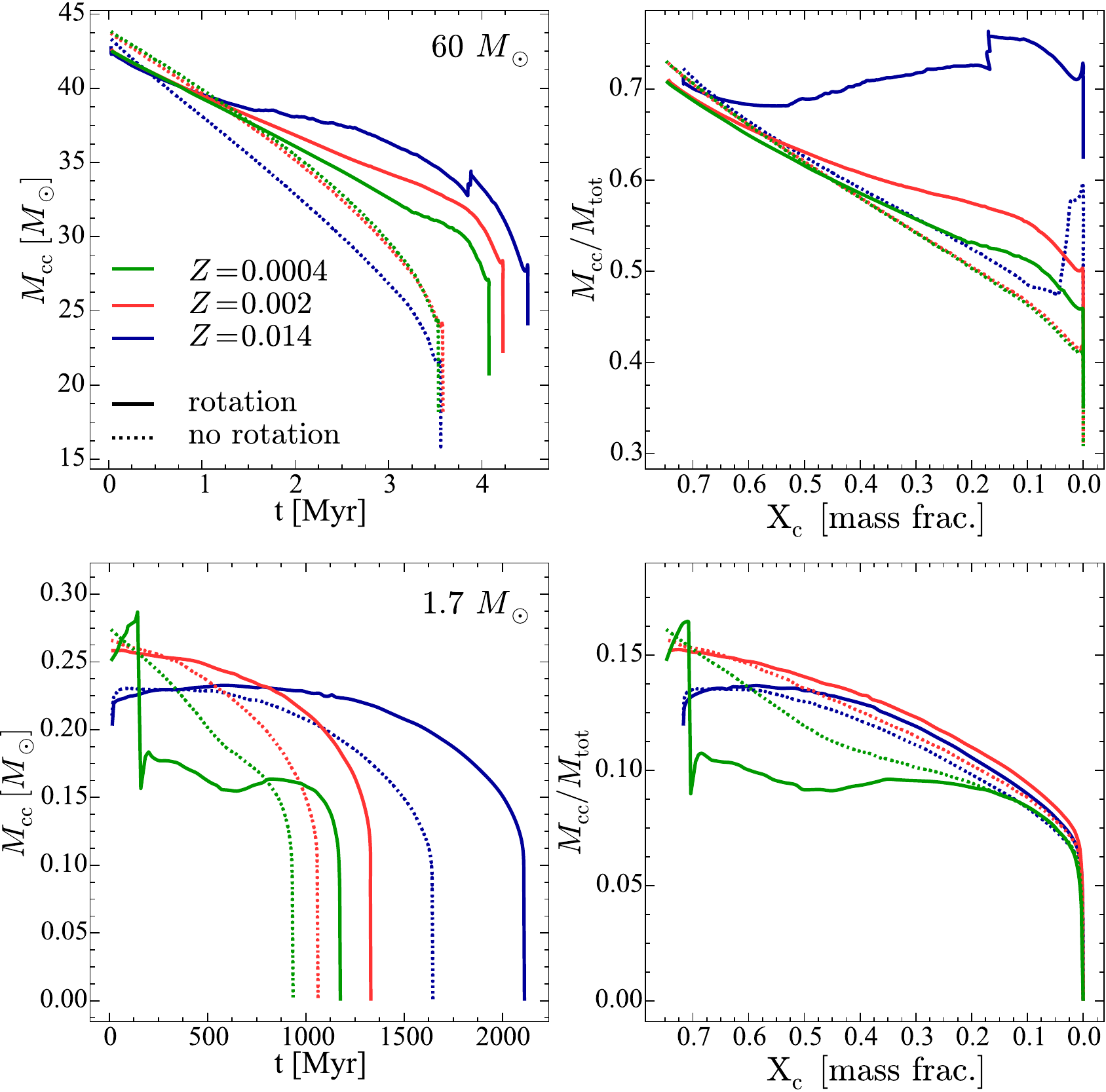}
\caption{Absolute value of the convective core mass as a function of age
during the core H-burning phase (left panels) and  fractional mass of the convective core as a
function of hydrogen mass fraction at the center (right panels). {\it Top panels :} 60 $M_\odot$ model.  {\it Bottom panels :} 1.7 $M_\odot$ model. 
\label{Fig:MccZ}}
\end{figure}

\subsection{Internal structure and lifetimes\label{Subsec:metaltime}}

Our models indicate that metallicity has a strong effect on the internal structure of stars, in particular when rotation is taken into account. Figure~\ref{Fig:MccZ} shows the mass of the convective core for models of four representative initial masses. The mass of the convective core is higher in rotating models compared to those without rotation, and the relative increase in the core mass between rotating and nonrotating models increases with metallicity. This effect also increases with initial mass, since the shear diffusion coefficient scales with radiative diffusion, which in turn increases with initial mass. The differences in internal structure between rotating and nonrotating models also affects their envelope properties, and ultimately changes the evolutionary tracks seen in the HR diagram (Fig.~\ref{Fig:HRDZ}).

\begin{figure}
\begin{center}
\includegraphics[width=0.95\columnwidth]{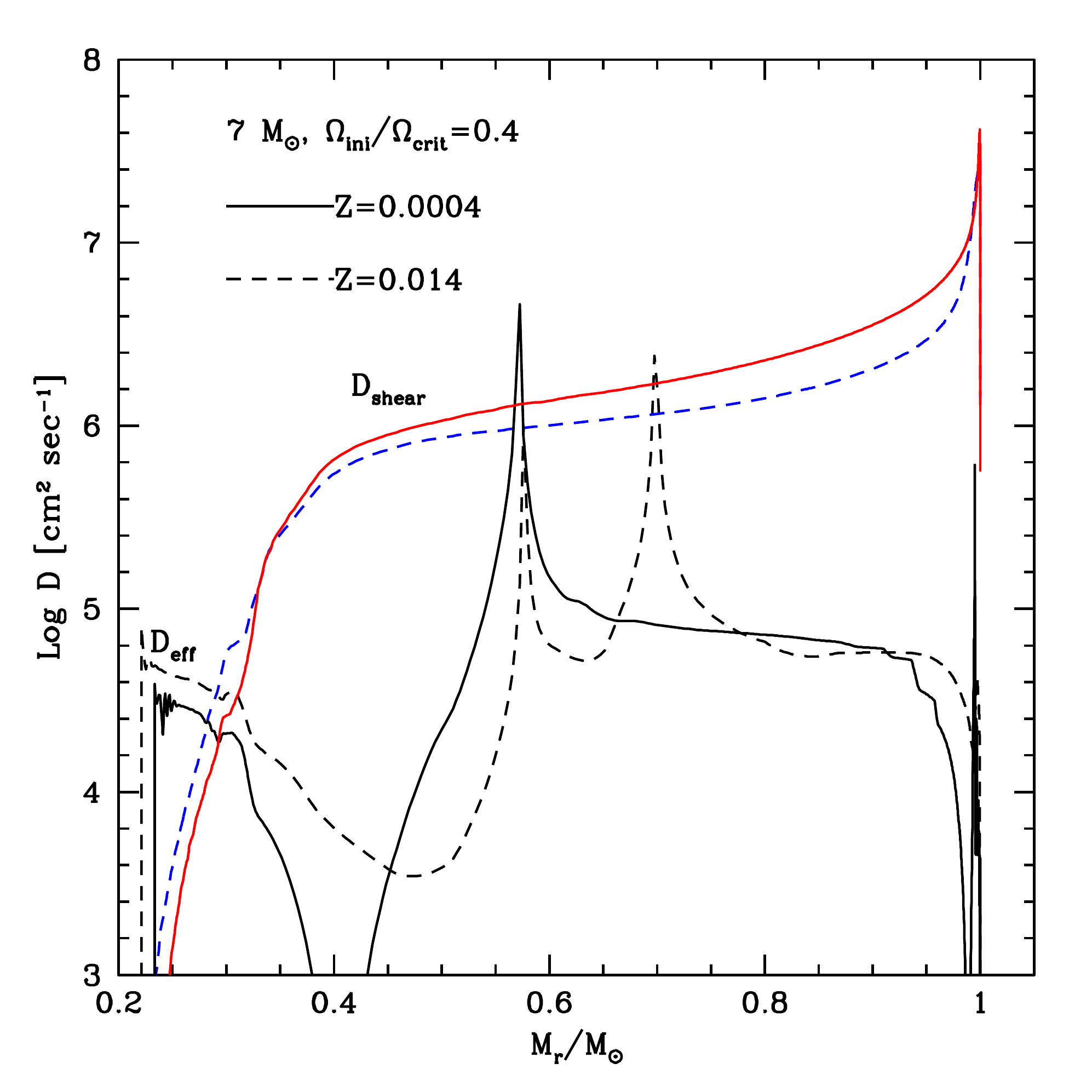}
\caption{Variation of the diffusion coefficients D$_{\rm shear}$ and D$_{\rm eff}$ as a function of the Lagrangian mass, normalized to the total mass of
rotating 7 $M_\odot$ models at two different metallicities. Both models have a mass fraction of hydrogen in their core equal to 0.4, i.e., they are both in the middle of the core H-burning phase.}
\label{D7}
\end{center}
\end{figure}

Why do we have larger increases of the convective core due to rotation at higher metallicities? Immediately above the convective core, shear mixing is strongly limited because of the inhibiting effects of the gradients of chemical composition. In this region, mixing is regulated by the net effect due to both meridional currents and a strong shear horizontal turbulence, which is encapsulated by the coefficient $D_{\rm eff}$ in our models \citep[see also Fig.~\ref{D7}]{CZ1992}.
Very schematically,  $D_{\rm eff}$ scales as $r U_r$, where $r$ is the distance to the center of the star and $U_r$ the radial component of the meridional current velocity. When the metallicity increases, stars become less compact and both $r$ and $U_r$ increase, since $U_r$ depends on the inverse of the density\footnote{Midway through the core H-burning phase ($X_c\sim 0.4$), the 7 $M_\odot$ model has 21\% of its mass (approximately the mass of the convective core) enclosed in a radius 23\% smaller in the Z=0.0004 model compared to the Z=0.014 model.} among other factors. This tends to make $D_{\rm eff}$ larger in stars that are more metal rich (see Fig.~\ref{D7})\footnote{At low metallicities, everything else being equal, the extension due to the overshooting will be smaller simply because mass is more concentrated.}.

%lifetimes as a function of metallicity
\begin{figure}
\centering \includegraphics[width=.89\columnwidth]{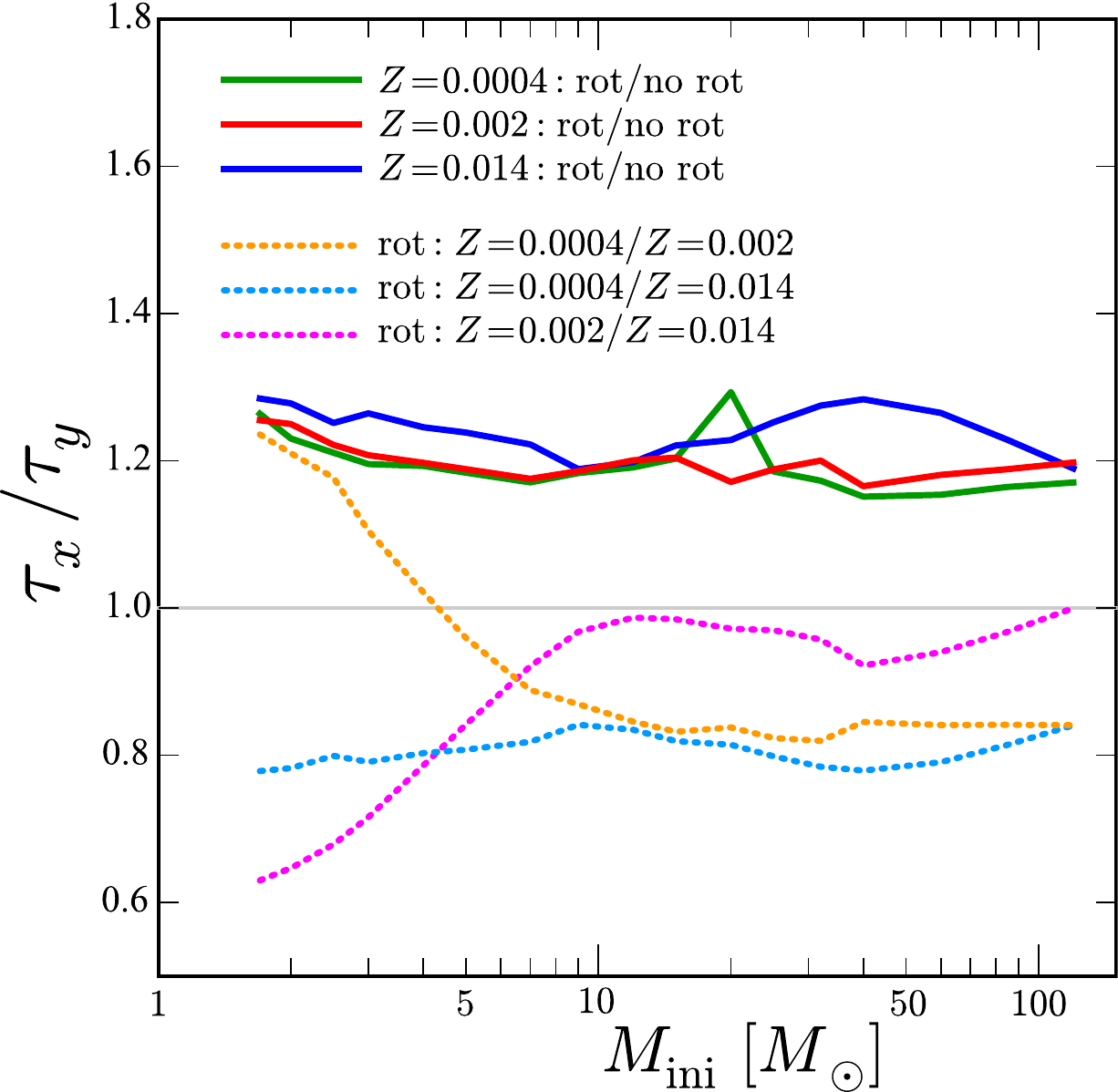}
\caption{ Ratio between the lifetimes of rotating and nonrotating Geneva stellar evolution models for different metallicity grids (solid lines). The dotted lines show the ratio between the lifetime of rotating models at two metallicities. The models at $Z=0.014$ are from \citetalias{Ekstrom2012a} and at $Z=0.002$ from \citetalias{Georgy2013a}.
\label{Fig:lifetimesZ}}
\end{figure}

Comparing the stellar lifetimes as a function of metallicity, rotating models have lifetimes increased by about 20\% with respect to the nonrotating models for the initial value of rotational velocity used in this paper. This increase is roughly independent of metallicity in the range Z=0.0004--0.014 (Fig.~\ref{Fig:lifetimesZ}). The lifetimes of the Z=0.0004 rotating models are about 80\% the lifetimes of the corresponding models at Z=0.014 in the entire metallicity range considered here (dotted blue line in Fig.~\ref{Fig:lifetimesZ}).

\begin{figure*}
\includegraphics[width=0.48\textwidth]{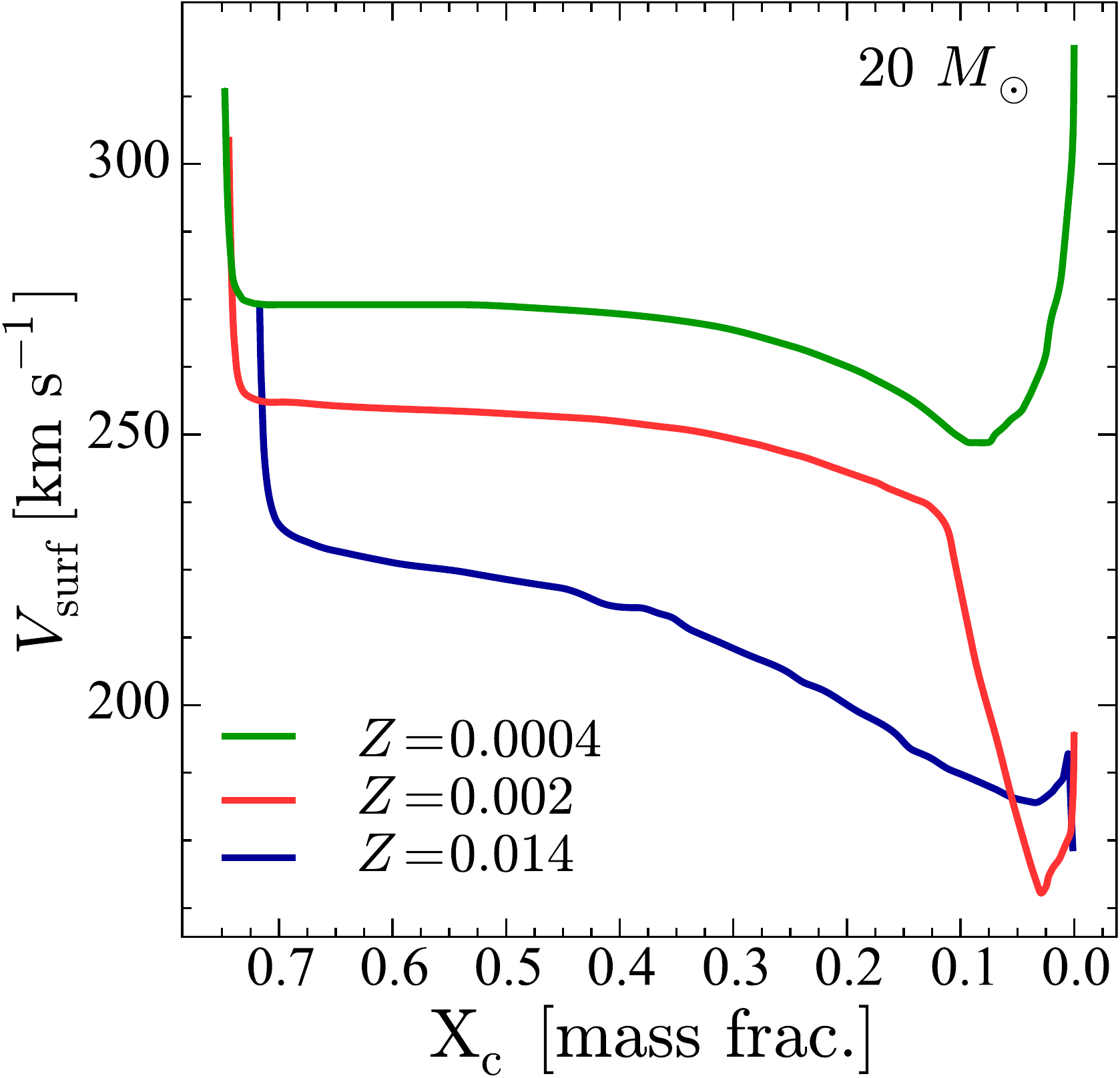}\hfill\includegraphics[width=0.48\textwidth]{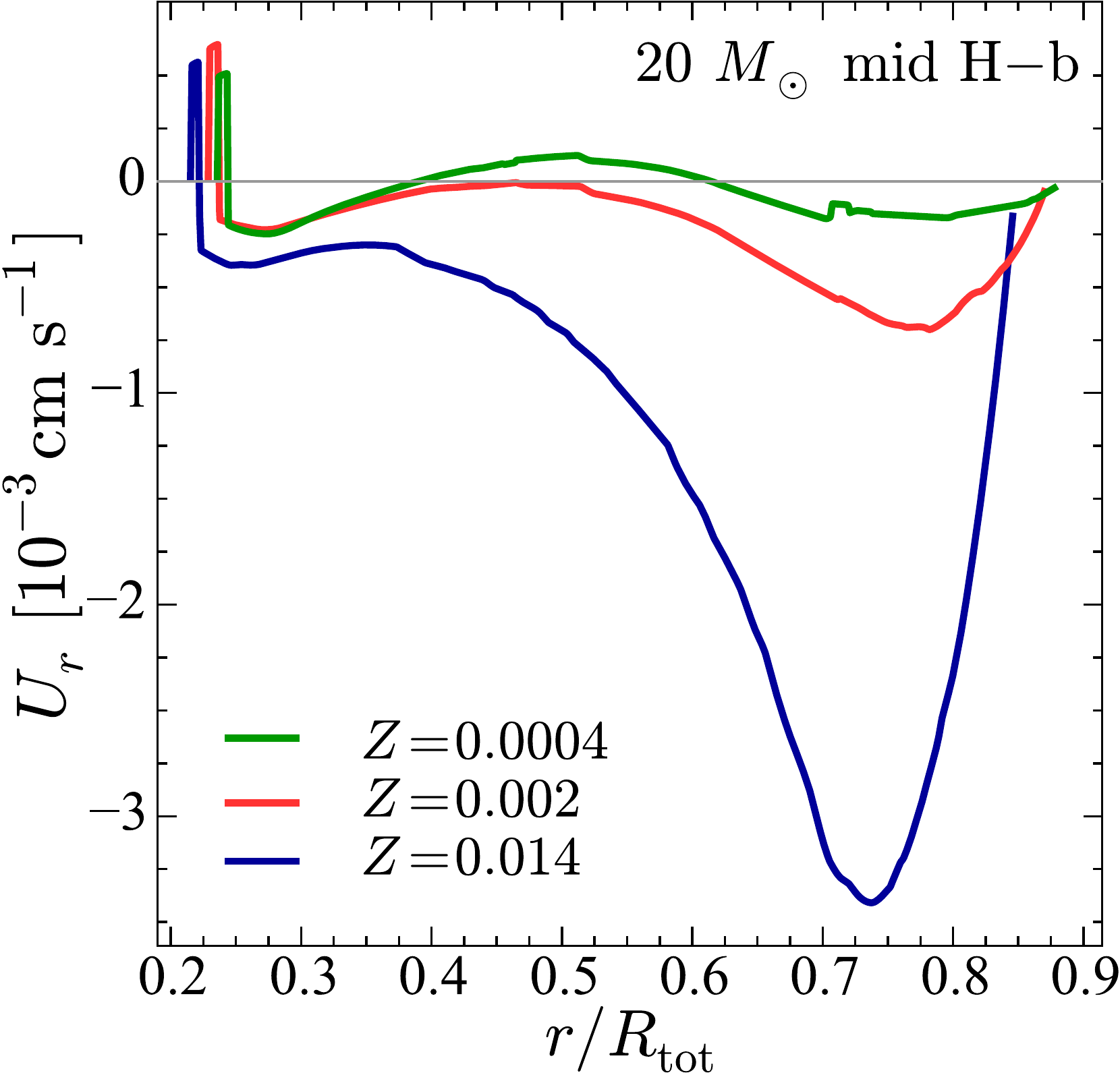}
\caption{\textsl{Left panel:} Surface velocity evolution during the MS for the $20~\msun$ models at various metallicities. \textsl{Right panel:} Radial component of the meridional circulation ($U_r$) inside the same models at the middle of the MS ($X{\rm c}\simeq0.35$). Negative values of $U_r$ mean that angular momentum is transported from the inner regions of the star to its outer parts, and a positive value corresponds to an angular momentum transport from the outer parts to the inner parts of the star.
\label{Fig:rotZ}}
\end{figure*}

Focusing specifically on rotating models, we identify different dependences of the stellar lifetimes with metallicity depending on the initial mass (Fig.~\ref{Fig:lifetimesZ}).
For initial masses lower than 8~$M_\odot$, the lifetime initially increases from $Z=0.014$ to $Z=0.002$, and then decreases from $Z=0.002$ to $Z=0.0004$. The magnitude of this nonmonotonic behavior is inversely proportional to the initial mass and depends on the complex interplay of opacity, rotation, and convection on stellar lifetimes in this mass regime. First, the opacities are strongly dependent on metallicity, making the star more luminous and with shorter lifetimes as metallicity decreases. Second, the effect of rotation on the hydrostatic structure tends to decrease the central temperature and thus increase the contribution of the pp chain to the total nuclear energy output, which tends to increase the lifetime. Third, rotation affects the evolution of the mass of the convective core. Since the diffusion coefficient $D_{\rm eff}$ diminishes for low metallicities, this tends to make the convective core less permeable to arrival of fresh fuel from the radiative zone and thus decreases the enhancement of the lifetime due to rotation.The final outcome of these counteracting effects cannot be predicted without numerical computations.

For initial masses above about 10 $M_\odot$, the lifetimes do not change substantially from Z=0.014 to 0.002, and decrease by about 20\% from Z=0.002 to Z=0.0004. In this initial mass regime, the dominant source of opacity is electron scattering, which is weakly affected by metallicity. However, we saw above that
the extension of the convective core both due to overshoot and to rotation will be diminished in low-metallicity models. These factors tend to reduce the lifetimes at lower Z. The exact amount of decrease is  difficult to obtain from simple principles since the lifetime will be an intricate interplay between the luminosity and the fuel available, and again numerical models such as those presented here are needed to properly estimate stellar lifetimes as a function of metallicity.

\subsection{Surface rotation velocities\label{Subsec:metalrotation}}

The impact of metallicity on surface rotational velocities is regulated by the effects of mass loss and angular momentum transport from the core to the surface. Here we investigate a model with an initial mass of 20~\msun\ at different metallicities, for which we present the evolution of the surface rotational velocity (Fig.~\ref{Fig:rotZ}, left panel) and the radial component of the meridional circulation velocity at the middle of the core H-burning phase (Fig.~\ref{Fig:rotZ}, right panel). 

The initial surface rotational velocities were chosen as 40\% of the critical velocity, making their absolute values higher as metallicity decreases because the stars are more compact. Initially, the models behave in a similar fashion and show a rapid decrease of $\vrot$ caused by the readjustment of the angular momentum structure, as the models were assumed to rotate as a solid body on the ZAMS.  The surface velocity of the Z=0.014 model decreases significantly as the star evolves while the models for Z=0.002 and Z=0.0004 show a roughly constant $\vrot$ until $X_c=0.30$, and then a minor decrease until $X_c=0.10$. This happens because of the impact of mass loss on angular momentum removal from surface layers, which is stronger at high $Z$.

Since the average densities are higher at low metallicity and $U_r$ varies with the inverse of the density in the outer layers, $U_r$ is lower at low metallicity, making the transport of angular momentum from the core to the surface less effective in metal-poor stars. However, because the removal of angular momentum at the surface is also less efficient at low metallicities, these two effects counterbalance each other during the majority of the core H-burning phase in the  Z=0.002 and Z=0.0004 models. The transport of angular momentum to the surface dominates over the angular momentum loss at the surface due to mass loss only when $X_c<0.10$ ($X_c<0.05)$ for the Z=0.0004 (Z=0.002) model. This leads to the spin up episode seen in these models at the end of core-H burning. 

This interplay is responsible for the outcome seen in the evolution of the surface velocity on the left panel of Fig.~\ref{Fig:rotZ}. At Z=0.014, the transport of angular momentum from the interior to the envelope, despite being stronger than at low metallicities, does not compensate for the losses of angular momentum at the surface and the increase in radial size of the star. At lower metallicities, the less efficient transport of angular momentum from the interior to the surface compensates for both the  increase in radius and the weak losses of angular momentum by the stellar winds.

\subsection{Surface abundances\label{Subsec:metalabund}}

\begin{figure}
\centering
\includegraphics[width=0.75\columnwidth]{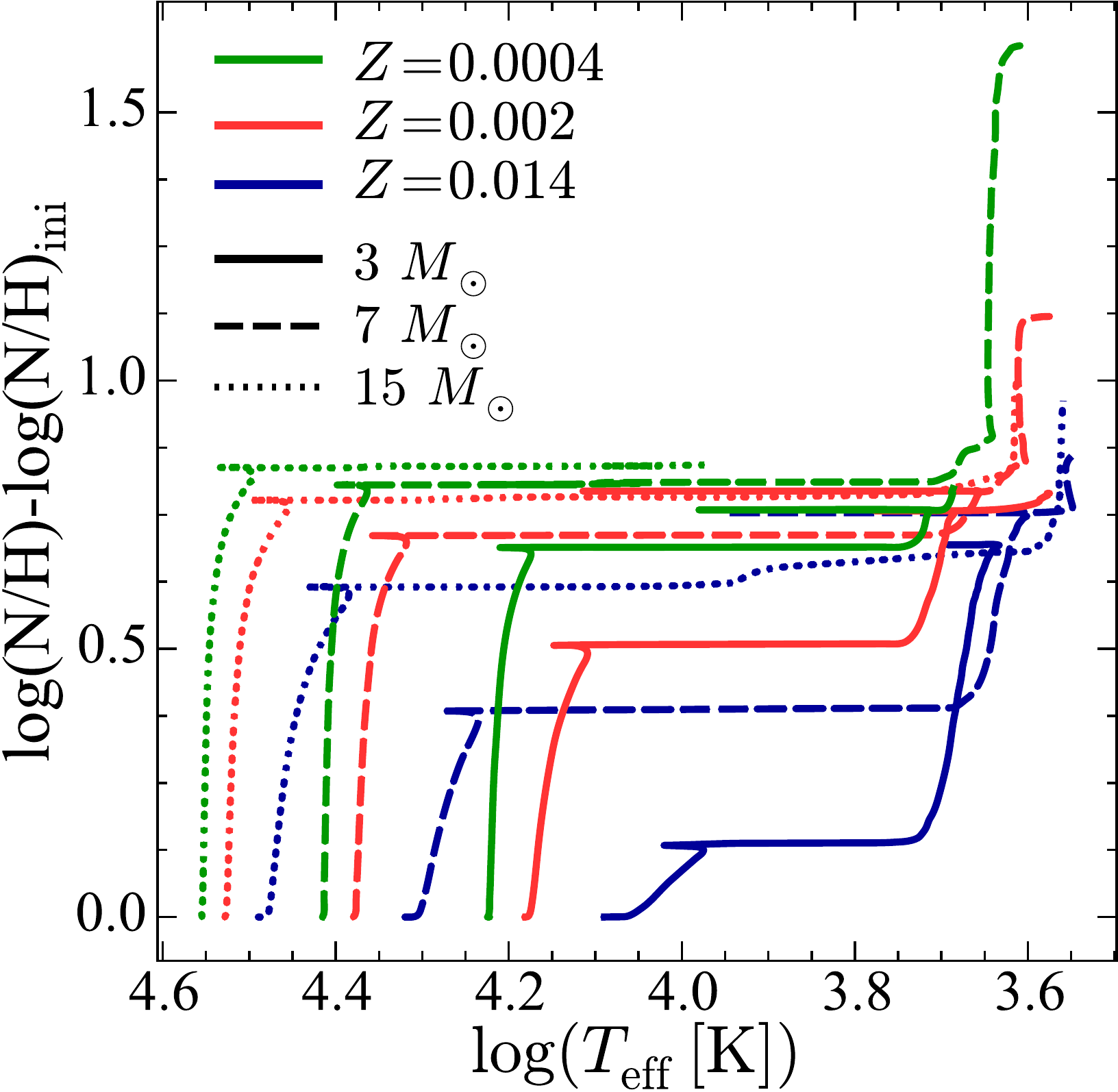}
\caption{Evolution of the relative enrichment $\log({\rm N/H})-\log({\rm N/H})_{\rm ini}$ as a function of \teff\  for the rotating $3$, $7$, and $15~\msun$ models at various metallicities.}
\label{Fig:NHZ}
\end{figure}

Figure~\ref{Fig:NHZ} shows the relative N enrichment at the surface of the rotating $3$, $7$, and $15\,M_\odot$ models at various metallicities. 
The relative N surface enrichment at the end of the MS phase, at a given mass, is larger at lower metallicities.
The N enrichment is stronger at low metallicity because stars at low metallicity are more compact, with stars at Z=0.0004 being about 20\% smaller than at Z=0.0014. Since the mixing timescale depends on $r^2/D$, where $D$ is the total diffusion coefficient, the smaller the radius $r$, the smaller the timescale. Moreover $D$ is also slightly larger at low metallicity because the gradient of $\Omega$ is slightly larger in more compact stars.

The increase in the relative N enrichment due to rotational mixing is larger for lower masses, with the $Z=0.0004$ models showing more enrichment than the $Z=0.014$ models by $0.55$, $0.42$, and $0.22$ dex for stars with $3$, $7$, and $15\,M_\odot$, respectively. The relative N enrichment of a Z=0.0004 3 $M_\odot$ at the end of the MS is larger than that of 15 $M_\odot$ star at Z=0.014. This occurs because meridional currents in the stars with higher initial masses and/or higher metallicities  are stronger and therefore more efficient in smoothing the gradients of $\Omega$ between the core and the envelope that drives the chemical mixing.

An observational consequence of the above fact is that at $Z=0.0004$, keeping constant the ratio between the initial rotation and the critical velocity, one expects surface enrichments due to rotation to be less dependent on the
initial mass than at higher metallicity. For example, the surface nitrogen abundance in the 3 $M_\odot$ at $Z=0.0004$, at the end of the MS phase is about 4.5 times the initial value, while the 15~\msun\ model shows an enhancement factor of 6.0 (cf. at $Z=0.014$, the N enhancements are  1.3 for the 3~$M_\odot$ model and 4.0 for the 15~$M_\odot$ model).

Comparing rotating and nonrotating models, we find that the N enhancement relative to its initial abundance is higher at low metallicity compared to high metallicity. The surface abundances reflect the efficiency of the diffusive transport in the radiative layer of a given element. Stronger surface N enrichment does not necessarily mean that globally the whole star is more mixed. Shear mixing is mainly responsible for mixing N in the radiative envelope. The stronger efficiency of shear mixing at low Z explains the stronger relative enrichment of N at the surface compared to higher-Z models.

Conversely, the evolutionary tracks on the HR diagram are less affected by rotation at lower metallicity than at higher metallicity. The impact of rotation on the HR tracks depends on the change of the convective core mass, which in turn depends on the diffusion coefficient $D_{\rm eff}$. The decrease of $D_{\rm eff}$ at low Z explains why the evolutionary tracks are relatively less affected by rotation compared to high metallicities.

It should be noted that the metallicity effects discussed in this subsection are strongly dependent on the prescription for the diffusion coefficients. \citet{meynet13a} investigates the impact of the different combinations of the diffusion coefficients used in the shellular rotating models, finding that the surface N abundances may vary significantly. Therefore, a quantitative comparison of our results with previous studies  (e.g., \citealt{SZ2015,Choi2016}) that employ different prescriptions for the diffusions coefficient is not straightforward. Still, we qualitatively compare our results to those studies in Sect. 5.

%=================================================================================
% METALLICITY EFFECTS
%=================================================================================

\subsection{Final fate of massive stars: Luminous blue variables, pair-instability supernovae, and massive black holes}

\begin{figure*}
\centering
\includegraphics[width=0.65\textwidth]{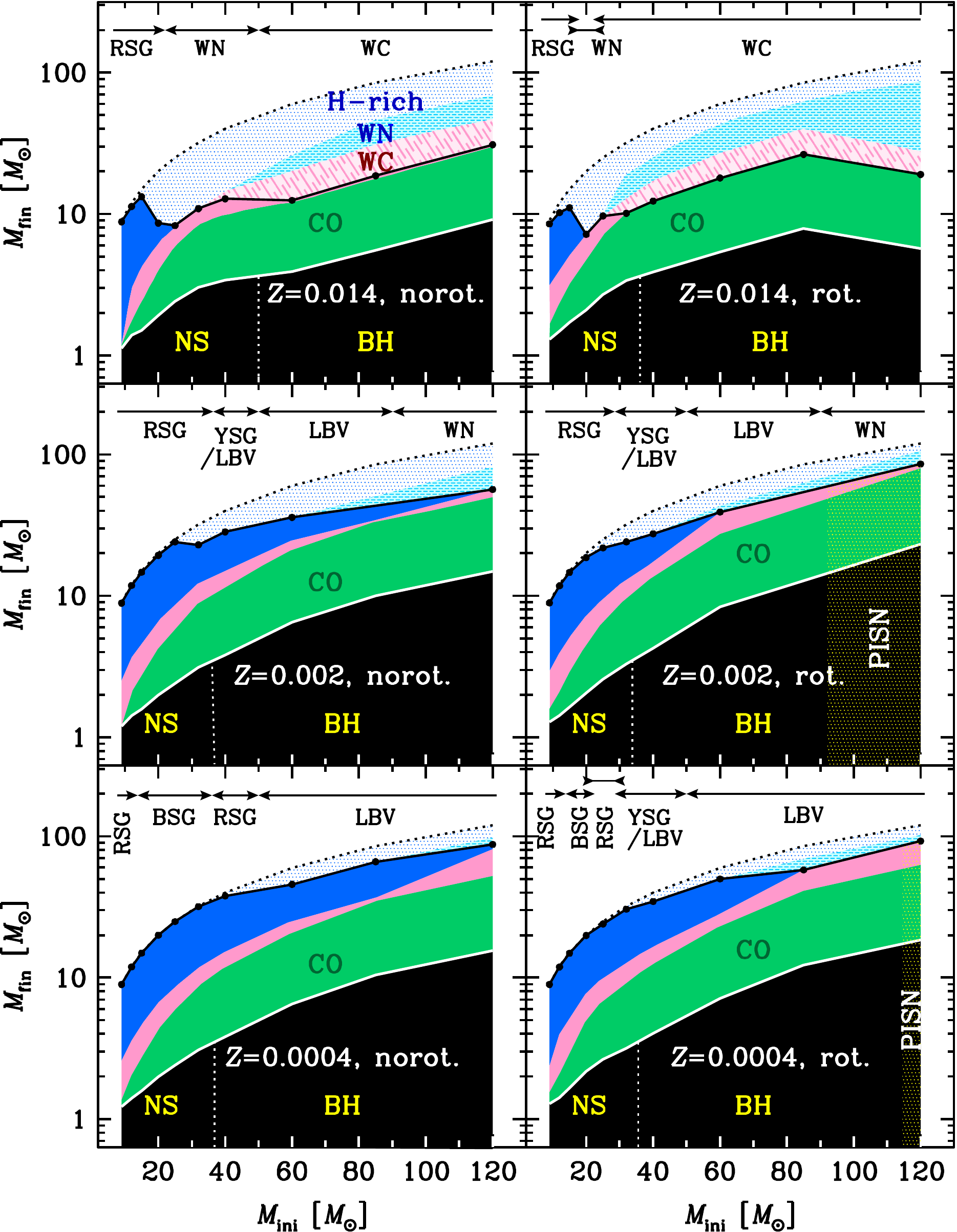}
\caption{Masses of the stellar remnants and under the form of SN and wind ejecta are shown as a function of the initial mass of stars for rotating and nonrotating models at three metallicities. The masses that remain locked into the BH or NS correspond to the solid black region. The labels at the top of each subpanel indicate the type of the core collapse progenitors. The yellow hatched zone in the lower right panel labeled PISN corresponds to the domain of masses that is predicted to produce PISNe. We also indicate the masses ejected at the time of the SNe (assuming a successful explosion for all masses) that are rich in carbon, oxygen, and other $\alpha$-elements (solid green regions labeled CO), or in H-burning processed material (solid pink regions), or with the same abundances as the initial one (solid blue regions). The hatched zones are the masses ejected by stellar winds, which are color coded in blue (H-rich), cyan (H-poor, labeled WN),  and pink (H-free, labeled WC). }
\label{Fig:end}
\end{figure*}

%final mass
Figure~\ref{Fig:end} shows the final fates of massive stars with and without rotation for the grid of models with metallicities $Z=0.0004$, $Z=0.002$, and $Z=0.014$. As expected, the final masses decrease as a function of metallicity because of higher mass loss from stellar winds at high metallicity \citep{Vink2001}.

%wind compared to ejecta
Let us compare the mass ejected by winds and the stellar mass at the final stage of the evolution. At solar metallicity, above an initial mass of $\sim50~M_\odot$ as much mass is ejected by stellar winds as in a possible SN explosion. Below an initial mass of 15 $M_\odot$, the contributions of the stellar winds are very modest for all metallicities; they are important, though not dominant, between 15 and 50 $M_\odot$. The limits depend slightly on rotation for the values explored here. For the range of metallicities studied in this paper, the wind contribution diminishes with decreasing $Z$ and never dominates at low Z. At $Z=0.0004$, stellar winds either have a composition similar to the initial abundance value or show CNO-processed elements (e.g., for the more massive models at advanced stages). This is in contrast with the solar metallicity models, which in some cases even show He-burning products in their winds.

%Massive Black Holes
Figure~\ref{Fig:end} shows the amount of mass ejected by SNe across the mass range studied in this paper, assuming that successful explosions happen for the 9-120~\msun\ range. To compute the ejecta mass, we follow the procedure described by \citet{georgy09}, which we briefly reiterate here. First, we compute the baryonic mass of the remnant using the relationship between the CO core mass and the remnant baryonic mass from \citet{maeder92}. The gravitational mass of the remnant is obtained with the relationship from \citet{hirschi05}. Then, the ejecta mass is computed assuming that all the mass from the edge of the remnant to the surface is ejected during the SN.

\begin{figure*}
\centering
\includegraphics[width=0.95\textwidth]{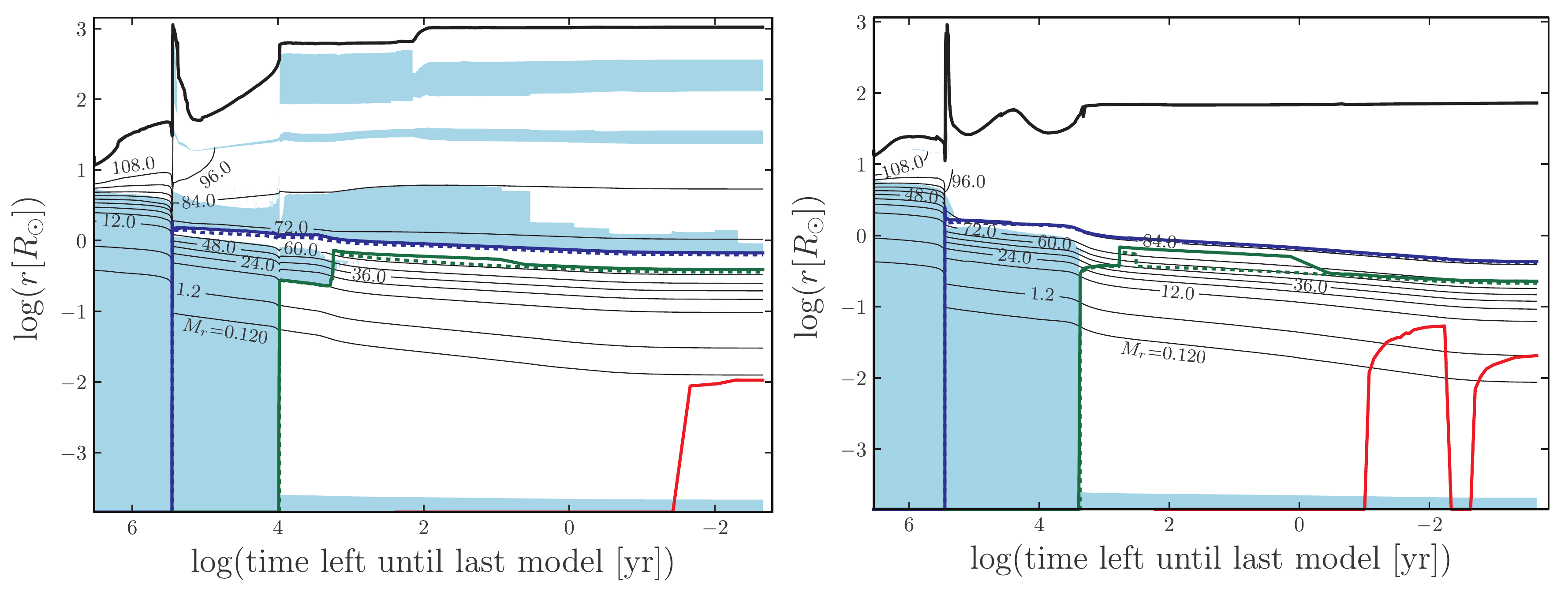}
\caption{ Kippenhahn diagrams for 120~\msun\ models at Z=0.0004 (left panel) and Z=0.002 (right panel), showing the evolution of the stellar structure in terms of the distance from the center as a function of time to core collapse. The black line at the top indicates the total radius of the star. The thin black lines followed by a label indicate the radial coordinate of different mass shells inside the star. The blue-shaded areas correspond to convective regions. The solid (dashed) lines correspond to the peak (10\%)  of the energy generation rate for H burning (blue), He burning (green), and C burning (red). }
\label{Fig:kip}
\end{figure*}

Since there is considerable debate on the mass ranges of stars that are able to produce a bright SN \citep[e.g.,][]{oconnor11,ugliano12,sukhbold14,sukhbold16,sukhbold18,pejcha15,ertl16,muller16,ebinger18}, we also consider the alternative scenario of failed explosions. Let us investigate the maximum black-hole mass that is formed at different metallicities for stars with initial mass in the range 9-120~\msun, assuming that all the mass is engulfed by the black hole, such that its mass would be equal to the final mass of the star. Under these assumptions, the maximum mass of the black hole that our models would produce at solar metallicity is 30 $M_\odot$ for a nonrotating star with initial mass of 120 $M_\odot$. At Z=0.002, we find $M_\mathrm{BH,max}=60-65 M_\odot$, depending on rotation, while the Z=0.0004 models yield even more massive black holes of up to 90~\msun. These values are substantially higher than the masses of the BHs that have been detected by the LIGO experiment \citep{LIGOII2016,LIGOI2016}.

%PISN
Because most massive low-metallicity rotating models do not lose a significant amount of mass as they evolve, they produce a significant CO core at the end of their lives and are within the expected theoretical range for pair-instability \citep{hw02,langer07} or pulsational pair-instability SNe \citep{chatzopoulos12,woosley17}. According to our models, at $Z=0.002,$ stars with initial masess larger than 85~\msun\ would explode as PISNe, with their progenitors being WR stars of the WN subtype or LBVs. At $Z=0.0004$, only the rotating $120~\msun$ model explodes as a PISN, having as direct progenitor possibly an LBV with high mass-loss rates and close to the Eddington limit (Fig.~\ref{Fig:Mdot}). Since LBVs have been detected in dwarf galaxies with metallicities similar to those analyzed here \citep{izotov09}, our models would imply that they are in the immediate stage before core collapse. There is also a possibility of producing a superluminous SN because the SN ejecta may interact with a dense CSM created by the progenitor, converting kinetic energy into radiative energy and boosting the observed light curve \citep[e.g.,][]{smith07,chevalier11}.

Surprisingly, the initial mass range of PISNe is reduced in Z=0.0004 models with respect to those with Z=0.002. This is a result of the large convective shells associated with the H-burning shell for the Z=0.0004 models. Figure \ref{Fig:kip} shows the time evolution of the internal structure (Kippenhahn diagrams) for the 120~\msun\ models at Z=0.0004 and Z=0.002. In the Z=0.0004 model, a large convective shell develops during the core-He burning phase, and the H-burning shell remains at a nearly constant mass coordinate inside the star. Indeed, the burning front which is located at the base of this convective shell is continuously replenished with H by convection, remaining at an approximately constant mass coordinate. This prevents the He and thus the CO core from growing in mass in the Z=0.0004 models as much as it does in the Z=0.002 model. This explains why the CO cores in the Z=0.0004 models are smaller than those at Z=0.002 despite the fact that the stellar winds are weaker at Z=0.0004 than at Z=0.002. As a result, the range of initial masses that produce massive-enough CO cores to fall into the PISN regime is smaller at Z=0.0004 than at Z=0.002. This process indicates that the treatment of convection, not only in the stellar core but in the burning shells, has a significant effect on the final fate of stars, as investigated in detail by \citet{GSM2014}.

\begin{figure*}
\centering
\includegraphics[width=0.95\columnwidth]{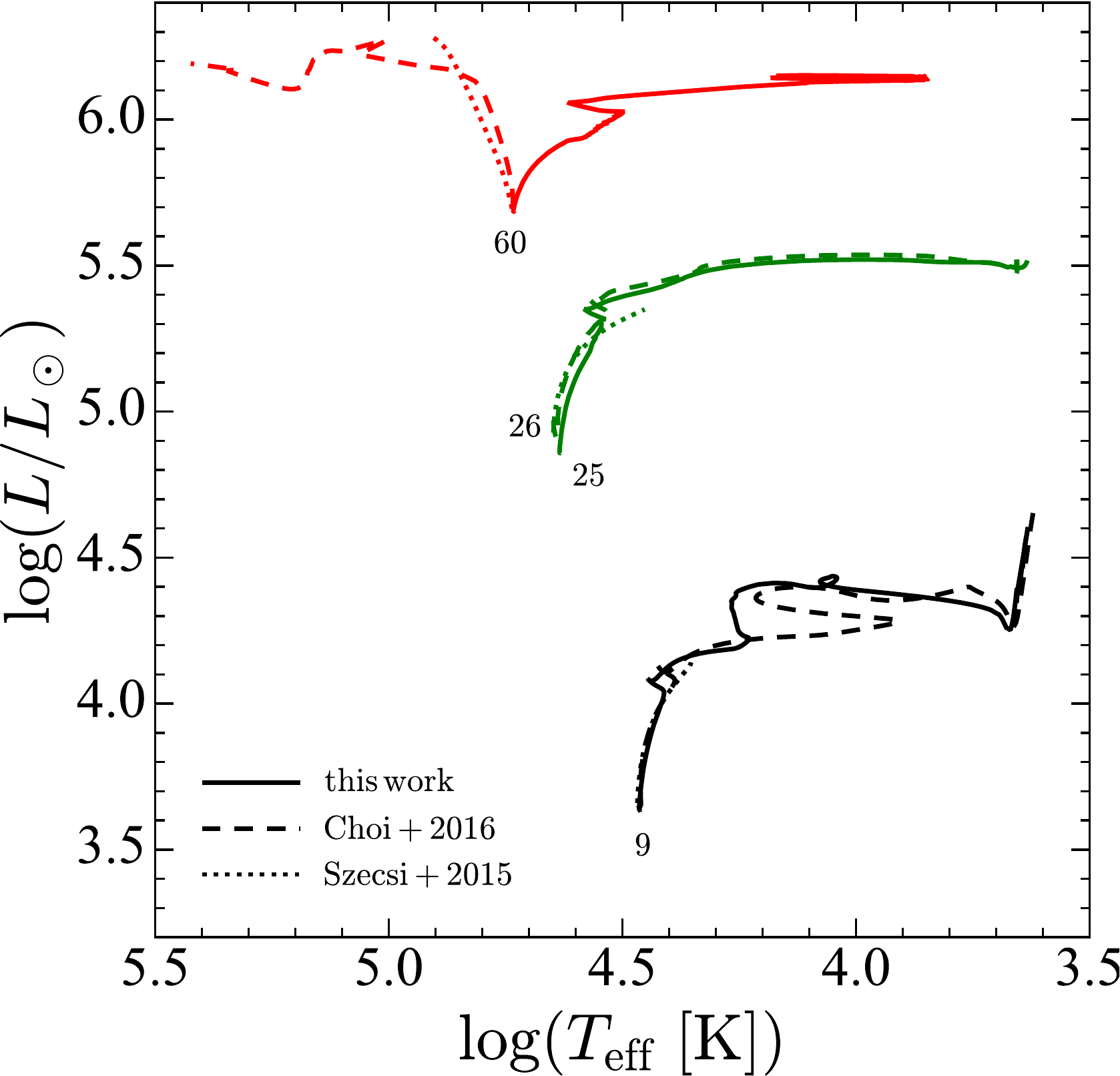}\includegraphics[width=0.95\columnwidth]{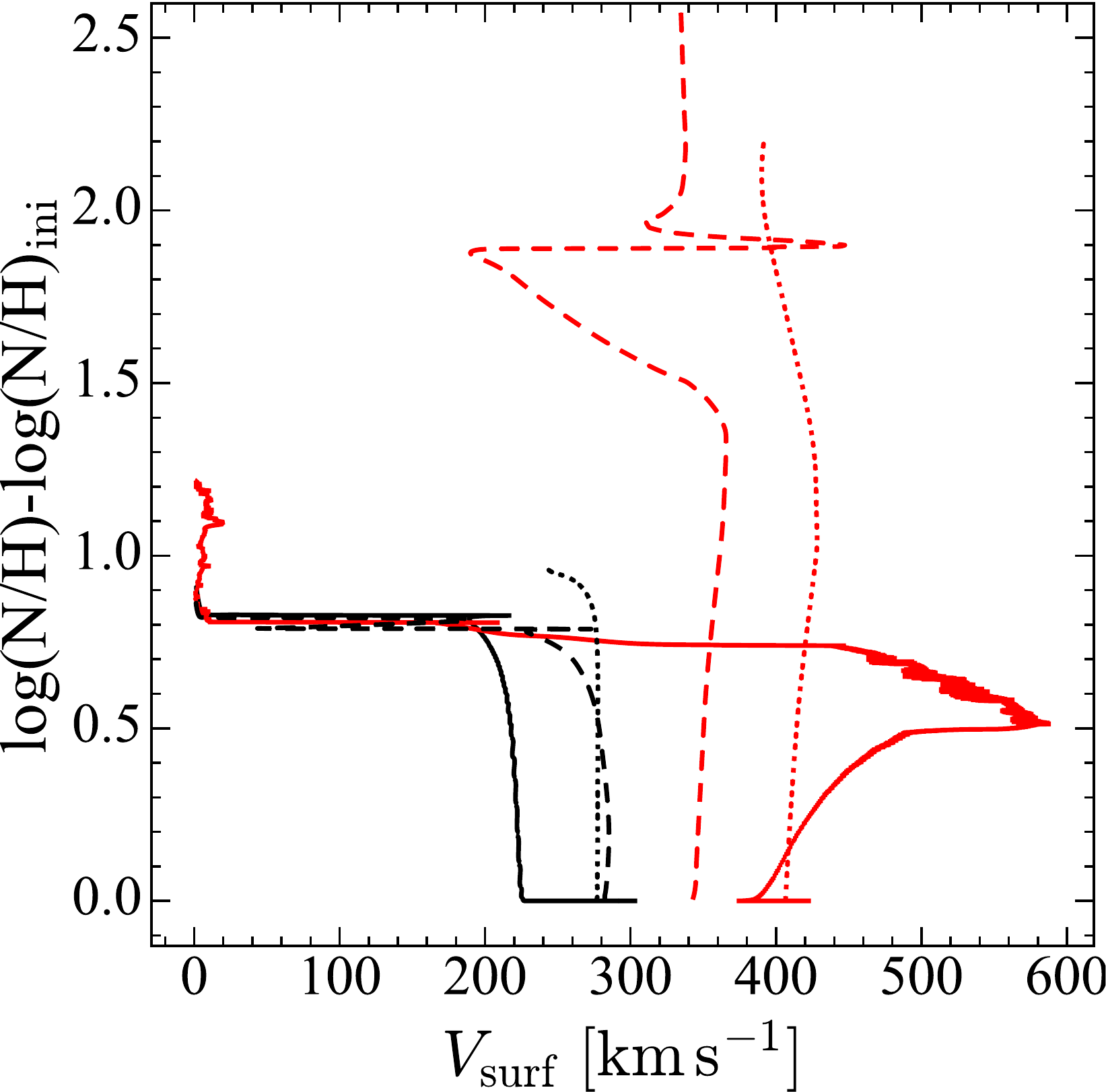}
\caption{{\it Left panel: }Comparisons of the evolutionary tracks for models with different initial masses in the theoretical HR diagram for a metallicity $Z$ around 0.00004, {\it i.e.,} corresponding to the metallicity of the galaxy I Zw 18. The code MESA models from Choi et al (2016) have [Fe/H]=-1.5 and no internal magnetic fields. The code STERN with an internal dynamo was used by Szecsi et al. (2015; we note that they computed a 26 $M_\odot$ model). {\it Right panel: } Evolution of the N surface enrichment as a function of the surface equatorial velocity for the 9 (black) and 60 $M_\odot$ rotating models (red). The meaning of the line styles are the same as in the left panel.  }
\label{Fig:HRDgenC}
\end{figure*}

% progenitor nature for different mass ranges
Figure~\ref{Fig:end} shows that the nature of NS progenitors is relatively diverse because of metallicity. Most of the stars producing a NS end their lifetimes as RSGs, with an upper mass limit strongly dependent on metallicity.  The upper initial limit for ending as an RSG is 22 $M_\odot$ for nonrotating models and 18 $M_\odot$ for rotating models at Z=0.01 \citep{Georgy2012,gmg13}, while at Z=0.002 it is 35 $M_\odot$ for the nonrotating models and 29 $M_\odot$ for the rotating models. These mass limits nearly coincide with the upper initial mass limit for the formation of NSs. However, for the rotating models in the mass range between 29 and 34 $M_\odot$ a SN event and MS formation may still occur with a yellow-blue supergiant, possibly an LBV star. At Z=0.0004, the NS progenitors are mostly blue supergiants in nonrotating models, while the rotating models also allow red supergiants and LBV/YHG progenitors.

\section{Comparison with previous studies and effects of different angular momentum transports}\label{Sec:comp}

In recent years, significant attention has been devoted to numerical stellar evolution models with metallicities similar to those presented in this paper ($\sim1/35\,\zsun$). In this section we compare our models with the most recent studies and discuss the qualitative differences. 

Figure~\ref{Fig:HRDgenC} shows the evolutionary tracks in the HR diagram and the N surface enrichments of rotating models computed by \citet{SZ2015} and \citet{Choi2016}, as well as those presented in this paper. Effects due to a slightly different initial mass and metallicity should be negligible, with most of the differences arising from the different implementations of angular momentum transport. In the model from \citet{SZ2015}, a very efficient angular momentum transport is included due to an internal magnetic field that has been amplified by a Tayler-Spruit dynamo \citep{Spruit2002}. In the MESA models from \citet{Choi2016}, a stronger angular momentum transport than that of our models
is considered. The stronger transport results from a different approach for computing the effects
of the meridional currents. These latter authors use a diffusive equation for describing the physics of angular momentum transport, while we compute this transport by solving an advective equation during the MS phase \citep{eggenberger08}. We note that an advective approach is in principle required since meridional transport is an advective -- rather than diffusive -- process by essence \citep{Zahn1992}. It should be noted though that, at least for the 20~\msun\ models at $Z=0.014$ and $Z=0.002$, the meridional currents carry angular momentum from the inner to the outer regions, that is, in the same direction that a diffusive process would do (Fig.~\ref{Fig:rotZ}, right panel).

However, the situation is less clear at $Z=0.0004$. In some cases, such as the 20~\msun\ model midway through the core H-burning phase (Fig.~\ref{Fig:rotZ}, right panel), the meridional currents carry angular momentum from the outer to the inner layers, i.e., in the opposite direction to that of a diffusive process. Therefore, it is particularly interesting to make a comparison at this metallicity. Another difference between our models and those from \citet{Choi2016} is the expressions that are used to describe $U_r$, with meridional currents appearing to be more efficient in the MESA than in the Geneva models. 

%describes 9, 25 (brief) and 60 Msun model
The evolutionary tracks for the rotating 9 and 25-26 $M_\odot$ models have some small differences, but all of them follow a similar qualitative evolution and finish as red supergiants. The case of the 60 $M_\odot$ model is very different. While  \citet{SZ2015} and \citet{Choi2016}  predict a homogeneous evolution, we predict a more classical behavior. In models with internal magnetic fields, such as the one from \citet{SZ2015}, there is almost solid-body rotation, and the meridional currents, which drive most of the chemical mixing, are very strong. In the model from \citet{Choi2016}, we suspect that even in the absence of an internal magnetic field, meridional currents are stronger and therefore also drive more mixing than in our models. One may conclude that to produce homogeneous evolution, `shellular' rotating models need higher initial surface rotational velocities (or larger initial content of angular momentum) than models with internal magnetic fields. Given an initial distribution of rotational velocities, our models without internal magnetic fields would predict a much smaller fraction of stars following a homogeneous evolution than the two other models at $Z=0.0004$.

%describes N enhancement
The N surface enrichments predicted by the different groups for the 9 and 60 $M_\odot$ models are shown in Fig.~\ref{Fig:HRDgenC}. The differences between them are very modest for the 9 $M_\odot$ models. This is partially because the diffusion coefficient responsible for the mixing of the elements has been calibrated in each family of models to reproduce the observed surface N enrichment of solar-metallicity MS B stars, in the mass range 9-15 $M_\odot$, and with a surface rotational velocity around the peak of the observed velocity distribution. The different 9 $M_\odot$ models still qualitatively agree at a lower metallicity of $Z=0.0004$. Conversely, the behavior of the N surface enrichment for the 60 $M_\odot$ models is quite different for the reasons described above, since the \citet{SZ2015}  and  \citet{Choi2016} models predict a homogeneous evolution and thus very strong N  enrichment at the surface.

%Surface velocities
Our standard 60 $M_\odot$ with rotation ($\vrot/\vcrit=0.40$, no internal magnetic fields) presents a surface equatorial velocity that increases significantly with time when evolution proceeds during the MS phase, while the  \citet{SZ2015}  and  \citet{Choi2016} models evolve with a nearly constant surface rotational velocity. At a given point, our model reaches critical velocity and mechanical mass loss slows down the surface layers by substantial amounts compared to the other family of models.

\begin{figure}
\includegraphics[width=0.99\columnwidth]{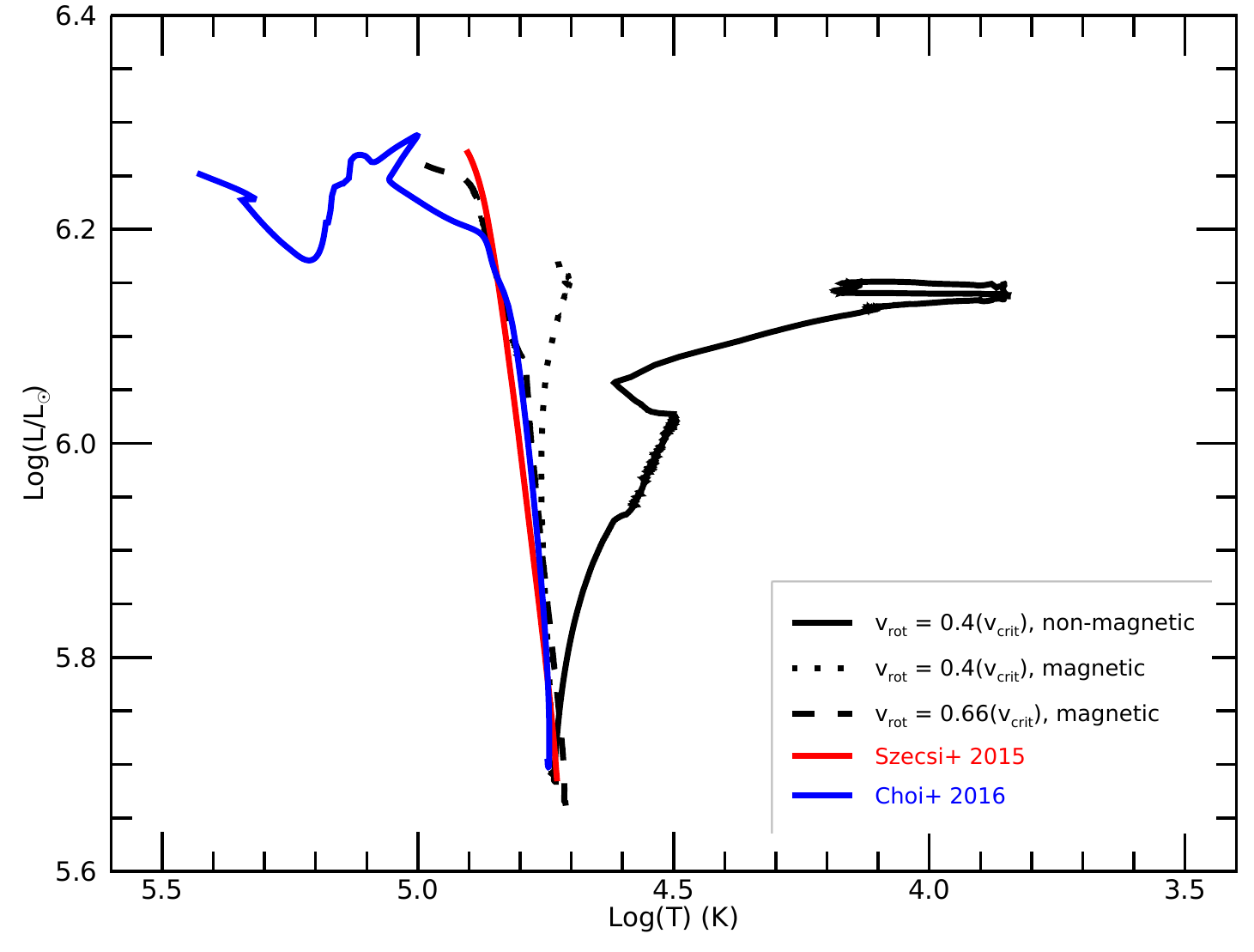}
\caption{Evolutionary tracks in the HR diagram for the $60~\msun$ models at $Z=0.0004$ computed by different groups. The black lines indicate the models computed in this paper, while the blue and red lines correspond to MESA \citep{Choi2016} and Bonn models \citep{SZ2015}.}
\label{Fig:var60}
\end{figure}

We computed two additional 60~\msun\ models including an internal magnetic field and different initial values of $\vrot/\vcrit$ (0.40 and 0.66) to explore their behavior under different assumptions about the angular momentum transport. These additional models include a Tayler-Spruit dynamo in a similar way as in \citet{Song2016}; we refer the reader to this latter publication for further details. Figure~\ref{Fig:var60} shows that, until the middle of the core H-burning phase, both our models with internal magnetic fields produce a similar evolution in the HR diagram to those of \citet{SZ2015}. From that point onwards, the model with internal magnetic field and $\vrot/\vcrit=0.40$ does not become as hot as the $\vrot/\vcrit=0.40$ model from \citet{SZ2015}. This type of evolution was referred to as ``transitional evolution'' by \citet{SZ2015}. To produce a similar evolution towards the hot side of the HR diagram, we need to increase the initial rotation to $\vrot/\vcrit=0.66$ with internal magnetic fields. This illustrates that rotational mixing is more efficient in the \citet{SZ2015} models than in our models.

%constraining angular momentum processes
Currently, it remains difficult to obtain strong constraints on the efficiency of angular momentum transport in massive stars. Asteroseismology is a promising approach, but it is challenging to apply it to massive stars. Some attempts  have been made to constrain the ratio between the core and envelope angular velocities for slowly rotating stars with initial masses around 9 $M_\odot$ \citep{Aerts2008}. In two cases, the core is rotating two to three times faster than the envelope. There is one case that is compatible with solid-body rotation. These values are well in the range of the values predicted by the current Geneva models \citep[see the discussion in][]{BEGRIDS2013}. However, asteroseismology of slowly rotating low-mass stars indicates that the internal transport of angular momentum is more efficient than predicted by meridional currents and shear instability. This is particularly the case for subgiant and red giant stars \citep{eggenberger12,eggenberger17,marques13,cantiello14}, but this is also suggested for low-mass MS stars \citep{kurtz14,saio15}. Upcoming facilities such as TESS \citep{ricker15} and PLATO \citep{rauer14} may allow us to obtain such observations for large samples of OB stars and to test the angular momentum transport theories in massive stars.

Another way of constraining the angular momentum transport is to study the evolution of the surface rotation for stars during the MS phase in a mass range where the mass-loss rates have a negligible impact on the angular momentum loss. In this respect, low-metallicity stars are ideal for this task as the line-driven winds become weak (see review by \citealt{puls08}). At solar metallicity, \citet{HG2006} showed that the Geneva models can provide a reasonable agreement with the evolution of the surface velocity of stars with initial masses in the range  9 -- 12 $M_\odot$.

The rotation rates of young pulsars are yet another possibility for constraining the evolution of the angular momentum transport in massive stars. \citet{HWS2005} showed that magnetic rotating models are in principle in better agreement with the rotation rate of young pulsars, since these models extract much more angular momentum from the core than those with differential rotation, and thus predict less angular momentum in the core at the pre-SN stage. Let us note that even the magnetic models do not seem to extract enough angular momentum from the core, since they fit only the lower values of the rotation period of young pulsars. In addition, it still remains to be seen if those objects suffer other instability processes at the time of the explosion and during the early phases of NS formation (e.g., magnetic braking), which may modify its angular momentum. All these additional complexities could make it harder to link the rotation of young pulsars and that of the pre-SN core of massive stars.

\section{Implications for IZw18 and other metal-poor dwarf galaxies}\label{Sec:izw18}

The blue-compact dwarf galaxy IZw18 is an ideal testbed for stellar evolution models at low metallicity for a number of reasons. First, it is one of the most metal poor star-forming galaxies in the nearby Universe. Its oxygen abundance is estimated to be about 1/35 to 1/50 the solar value \citep[see {\it e.g.,}][]{Izo1999,kehrig16}. This low level of oxygen enrichment indicates that for a long time the star-formation rate in this galaxy was quite low.  Second, IZw18 shows regions of star formation in its north-western (NW) and south-eastern parts (SE). These strong starbursts, together with the very low metal content, make this galaxy similar to a primeval galaxy that would undergo its first important star-formation episode \citep{Izo2004}. These authors did not find any evidence for the presence of stars from the red giant branch, while their observations were deep enough to have detected stars at the tip of the red giant branch if they were present. The most evolved stars have therefore an age inferior to 500 Myr. 

Third, regions of this galaxy show evidence for strong sources of ionizing photons \citep{Kehrig2015, Kehrig2017}. Some authors have proposed that such regions might contain Pop III stars that were formed recently from a relic pocket of material that would have retained the composition inherited from the Big Bang nucleosynthesis \citep{Heap2015}. This view is somewhat supported by the discovery that the chemical composition of the HII regions is different from that of the HI regions \citep{Lebouteiller2013}. On the other hand, X-ray binaries may also be important sources of ionizing photons \citep{Lebouteiller2017}. Fourth, IZw18 shows a very low content of dust \citep{Herrera2012}. This is in stark contrast with SBS 0335-052, another low-metallicity starburst galaxy \citep{Hunt2014}. Fifth, the N/O ratio in the interstellar medium of this galaxy is relatively high \citep{Aloisi2003}, and the galaxy contains Cepheids that have been used to estimate its distance \citep{Contreras2011, Aloisi2007}.  Finally, the NW region of IZw18 shows broad spectral lines compatible with the presence of WR stars of the WN and WC subtypes \citet{Izo1997} \citep[see also][]{Legrand1997, Brown2002}.  The relatively high number of WR stars is surprising, since WR stars are in general rare in metal-poor regions \citep{Crowther2006}. According to \citet{Izo1997}, the ratio of the number of WR to that of O-type stars is around 0.02. 

A quantitative comparison of predictions based on stellar evolution models and observations rely on population synthesis coupled with stellar atmosphere models, which we will defer to future work. Here we discuss the main impacts of our results on interpreting galaxies such as IZw18.

Concerning WR populations, our models do not predict the existence of a significant number of classical WR stars in \izw\ because of the expected weakness of stellar winds at low $Z$. According to our models, massive stars become relatively cool in the post H-core burning phases, would likely appear as LBVs, and emit a small number of H and He ionizing photons. They would also have narrow lines in their spectrum, and it would be unlikely that they would be responsible for the broad WR features observed in \izw. This would support the idea that the WR stars observed in \izw\ \citep{Izo1997} are not produced by wind mass loss from a single massive star with initial mass up to 120~\msun, but from homogeneous evolution \citep{SZ2015} and/or binary mass transfer.  We note that our models do not produce homogeneous evolution for the initial rotational velocities considered here and without considering internal magnetic fields. There is also the possibility that the WR stars in \izw\ could originate from very massive stars with initial mass greater than those considered here (120~\msun). We note that  \citet{graefener15} found that very massive MS stars could produce narrow \ion{He}{ii} emission lines down to $Z\simeq0.0001$ and explain the strength of that emission line in high-redshift galaxies. However, very massive MS stars would in most likelihood not produce the classical WR signatures seen in \izw.

In the context of binary models, \citet{goetberg18} found that the mass stripping by Roche-lobe overflow becomes inefficient at low metallicities comparable to that of \izw, and would not produce a typical WR spectrum for primary stars with $\mini$ up to $\sim18~\msun$. More massive binaries seem to be required to produce WR stars similar to those seen in \izw, with their nature and evolutionary histories still being an open question. 

Finally, it is interesting to note that the high N/O from \izw\ may support the primary nitrogen production in massive stars. This is because a high N/O in a metal-poor galaxy would require N and O to be ejected into the interstellar medium at similar times during galaxy evolution \citep{matteucci86}. Since O is produced mainly by massive stars, a high N/O would imply that N has a primary origin in massive stars as well. While our current models may produce some primary N, it remains to be seen whether it is enough to reproduce the observations. In addition, the N/O in the interstellar medium may result from previous populations at still lower metallicities than the current value for \izw\ which are more efficient producers of primary N. In this case, the previous stellar generations would have had to pollute the interstellar medium via winds and faint SN events to enrich the gas in N but not much in Fe \citep{Meynet2006, Choplin2017}.

\section{Concluding remarks and observable consequences}\label{Sec:conc}

Here we present a grid of rotating and nonrotating stellar evolution models computed with the Geneva code, for the initial mass range 1.7--120~\msun\ and metallicity $Z=0.0004$. Below we summarize our main findings, which are applicable to stars evolving without strong interactions with a companion star, either because they are isolated or because the orbital separation is too large to strongly affect the evolution in the corresponding stages described below.

\begin{enumerate}
\item Stars show stronger relative surface enrichments in nitrogen at Z=0.0004 than at Z=0.002 and 0.014 (considering
a fixed initial mass, a fixed initial rotation and a similar evolutionary stage).
\item On average, stars are rotating faster at lower metallicities due to the fact that they are more compact and
due also to the weakening of line-driven stellar winds.
\item Stars with $\mini > 5-6~\msun$ spend most of the core He-burning lifetime as blue supergiants. Therefore, one expects a large blue-to-red-supergiant ratio in cases where single star evolution dominates the populations.
\item Cepheids should be relatively rare since the instability strip is,
in general,  only crossed after the end of the core He-burning phase. 
\item Cepheids always evolve in such a way that their pulsation period increases with age.
\item All the Cepheids originating from the present rotating models show surface He and N enrichments.
\item Very few red supergiants are expected at this low metallicity since the stars are predicted to enter into that phase only after the end of the core helium burning phase.
\item Very few WR stars, if any, are predicted by our models. Faster rotation or higher initial masses than those considered here might produce more WR stars, although they could be rare. Our models with internal magnetic fields produce chemically homogenous stars that would lead to WR signatures, confirming the results from \citet{SZ2015}. Interacting binaries could also be an efficient channel for producing WR stars at these low metallicities \citep[e.g.,][]{goetberg17}.
\item At core collapse, massive stars are either red supergiants, yellow-blue supergiants, or luminous blue supergiants. Some of these stars
possibly end their lives as LBVs and may produce a pair of instability SNe and/or a superluminous SN event.
\item The angular momentum in the core at the core collapse is typically at a level that may impact the SN explosion, or the production of an accretion disk around a black hole if the explosion is unsuccessful. However, we caution that the results discussed here do not include internal magnetic fields.
\item While some of the nitrogen produced is of primary origin (i.e., not from the initial CNO composition) because of rotational mixing, the secondary production of nitrogen becomes significant at $Z = 0.0004$ and for many masses dominates the total nitrogen production, even in rotating models. The metallicity of \izw\ thus corresponds to the transition from primary to secondary production as the dominant channel of nitrogen production in rotating stars.
\end{enumerate}

%=================================================================================
% ACKNOWLEDGEMENTS
%=================================================================================
\begin{acknowledgements}
We thank the referee, Dr Dorottya Sz{\'e}csi, for a detailed review of the paper and the many useful suggestions. JHG acknowledges support from the Irish Research Council New Foundations Award 206086.14414 `Physics of Supernovae and Stars'. GM is supported by the
Swiss National Science Foundation (project Interacting Stars, number 200020-172505). RH acknowledges support from the World Premier International Research Center Initiative (WPI Initiative), MEXT, Japan and from the ChETEC COST Action (CA16117), supported by COST (European Cooperation in Science and Technology). The research leading to these results has received funding from the European Research Council under the European Union's Seventh Framework Programme (FP 2007-2013)/ERC Grant Agreement No. 306901. LJM and EJF are supported by Irish Research Council Postgraduate Fellowships, and IB by a Trinity College Dublin Postgraduate Award. We appreciate comments on the original manuscript from Y. Izotov.

\end{acknowledgements}

%=================================================================================
%: BIBLIOGRAPHY
%=================================================================================
\bibliographystyle{aa}

\end{document}